\documentclass[12pt,technote,onecolumn]{IEEEtran}
\usepackage{cite}

\usepackage{multirow}

\ifCLASSINFOpdf
  \usepackage[pdftex]{graphicx}
\else
  \usepackage[dvips]{graphicx}
\fi
\usepackage[cmex10]{amsmath}
\usepackage{amssymb}

\newtheorem{theorem}{Theorem}
\newtheorem{lemma}{Lemma}
\newtheorem{assumption}{Assumption}

\newtheorem{definition}{Definition}
\newtheorem{example}{Example}

\newtheorem{remark}{Remark}

\newcommand{\C}{\mathcal{C}}

\usepackage{array}

\ifCLASSOPTIONcompsoc
  \usepackage[caption=false,font=normalsize,labelfont=sf,textfont=sf]{subfig}
\else
  \usepackage[caption=false,font=footnotesize]{subfig}
\fi
\usepackage{url}

\begin{document}
%
\title{Distributionally Robust Counterpart in Markov Decision Processes}
%
%
%

\author{Pengqian Yu, 
        Huan Xu
\thanks{This work is partially supported by the Ministry of Education of Singapore through AcRF Tier Two grant R-265-000-443-112.}
\thanks{P. Yu is with the Department of Mechanical Engineering, National University of Singapore, 9 Engineering Drive 1, Singapore 117575, Singapore  (e-mail: yupengqian@nus.edu.sg).}
\thanks{H. Xu is with the Department of Mechanical Engineering, National University of Singapore, 9 Engineering Drive 1, Singapore 117575, Singapore (e-mail: mpexuh@nus.edu.sg, Tel: (+65) 6516 4094).}}

\maketitle

\begin{abstract}
This paper studies Markov Decision Processes under parameter uncertainty. We adapt the distributionally robust optimization framework, and assume that the uncertain parameters are random variables following an unknown distribution, and seeks the strategy which maximizes the expected performance under the most adversarial distribution. In particular, we generalize previous study \cite{xu2012distributionally} which concentrates on distribution sets with very special structure to much more generic class of distribution sets, and show that the optimal strategy can be obtained efficiently under mild technical condition. This significantly extends the applicability of distributionally robust MDP to incorporate probabilistic information of uncertainty in a more flexible way.
\end{abstract}


%
\IEEEpeerreviewmaketitle

\section{Introduction}
%
%
%
%
\IEEEPARstart{M}{arkov} Decision Processes (MDPs) are widely used tools to model stochastic sequential decision making problems (e.g., \cite{puterman2009markov, bertsekas1996neuro, barto1998reinforcement}). A strategy that achieves maximal expected accumulated reward is considered optimal. However, in practice, the  transition probabilities and reward parameters are typically estimated from finite and possibly noisy data, which often deviate from their true values. Such deviation, called ``parameter uncertainty'', can cause the performance of the optimal policies to degrade significantly (see experiments in~\cite{mannor2007bias}).

Inspired by the ``robust optimization'' framework in mathematical programming (e.g., \cite{soyster1973technical, ben1999robust, bertsimas2004price, ben2009robust}), many efforts have been made to alleviate the effect of parameter uncertainty in MDPs (e.g., \cite{xu2012distributionally, nilim2005robust, iyengar2005robust, delage2010percentile, white1994markov, bagnell2001solving, epstein2007learning}).
Most previous study (e.g., \cite{nilim2005robust, iyengar2005robust, bagnell2001solving, epstein2007learning,wiesemann2013robust}) focuses on the ``robust MDP'' which treats the uncertain parameter as a fixed yet unknown element of a given ``uncertainty set'', and aims to find the strategy that achieves best performance under the worst parameter. This set-inclusive formulation of uncertainty can be conservative as it cannot incorporate probabilistic information of the uncertainty that is often available in practice (e.g., \cite{delage2010percentile,xu2006robustness}). To overcome this, \cite{xu2012distributionally} proposed the {\em distributionally robust MDP} approach, which can incorporates certain kind of probabilistic information of the uncertainty. More specifically, this  approach treats the uncertain parameters as a random variable {\em following an unknown distribution}, while the distribution is known to belong to a set of distributions, called the ``ambiguity set'', and the goal is to seek  a strategy that archives the maximum expected performance under the most adversarial distribution of the uncertain parameters. Indeed, this approach is the multi-stage counter-part of the distributionally robust optimization (e.g., \cite{xu2012distributionally,wiesemann2013distributionally}) which considers the following: Given a utility function $u(x,\xi)$ where $x\in \mathcal{X}$ is the optimizing variable and $\xi$ is the unknown parameter, distributionally robust optimization solves $\max_{x\in \mathcal{X}}[\inf_{\mu \in \C}\mathbb{E}_{\xi\sim\mu}u(x,\xi)]$, where $\C$ is a-priori known set of distributions.

We highlight our contributions by comparing with \cite{xu2012distributionally}. In \cite{xu2012distributionally}, the state-wise ambiguity set is restricted to the following form:
$\tilde{\C}_s=\{\mu_s|\mu_s\left(O_s^i\right)\geq \underline{\alpha}_s^i \quad \forall i=1,\dots, n_s\},$
where $\underline{\alpha}_s^i\leq \underline{\alpha}_s^j$ and $O_s^i$ is a proper set of uncertain parameters with a ``nested-set'' structure, i.e., satisfying $O_s^i\subseteq O_s^j$, for all $i<j$  (see Fig. \ref{figure1(a)}). This setup can effectively model distributions with a single mode (such as a Gaussian distribution),  but less so when modeling multi-mode distributions such as a mixture Gaussian distribution. Moreover, other probabilistic information such as mean, variance etc can not be incorporated. Thus, in this paper, we extend the distributionally robust MDP approach to handle ambiguity sets with more general structures. In particular, we consider a class of ambiguity sets first proposed in \cite{wiesemann2013distributionally} as a unifying framework for modeling and solving distributionally robust single-stage optimization problems, and embed them into distributionally robust MDPs setup. These ambiguity sets are considerably more general: they are characterized by a class of $O_s^i$ which can either be nested or disjoint (as shown in Fig. \ref{figure1(b)}), and moreover, additional linear constraints are allowed to define the ambiguity set, which can be used to incorporate probabilistic information such as mean, covariance or other variation measures. We show that, under this more general class of ambiguity set, the resulting distributionally robust MDPs remain tractable under mild technical conditions, and often outperforms previous methods partly due to the fact that it can model uncertainty in a more flexible way.

The rest of the paper is organized as follows. Section \ref{sec2} provides some background of uncertain MDPs and presents our problem setup and necessary assumptions. Finite horizon and discounted reward infinite horizon distributionally ambiguous MDPs are discussed in Section \ref{sec3} and Section \ref{sec4}, respectively. We are particularly interested in the solution approach, i.e., how to solve the distributionally ambiguous MDPs. We present simulation results on a machine replacement problem and a path planning problem, both with parameter uncertainty in Section \ref{sec5}, which shows that the proposed approach   outperforms the nominal, the robust, and the distributionally robust approach proposed by \cite{xu2012distributionally}. We conclude the paper in Section \ref{sec6}. As this work generalizes~\cite{xu2012distributionally}, we  reuse some of its results, which will be indicated when we encounter in the corresponding sections. 

\section{Preliminaries}\label{sec2}
Throughout the paper, we use capital letters to denote matrices, and bold face letters to denote column vectors. Row vectors are represented as the transpose of column vectors. We use $\mathbf{1}$ to denote the indicator function, and use $\mathbf{e}_i(m)$ to denote the $i^{th}$ elementary vector of length $m$. $\mathbb{R}^n_{+}$ denotes the set of nonnegative real $n$-vectors, and $\textbf{tr}(A)$ denotes the trace of square matrix $A$. If $\C$ is the set of joint probability distribution of three random vectors $\mathbf{a},\mathbf{b}$ and $\mathbf{c}$, then $\prod_{(\mathbf{a},\mathbf{b})}\C$ denotes the set of marginal distribution of $(\mathbf{a},\mathbf{b})$. We use $\oplus$ to represent mixture distribution: given two probability distribution $\mathcal{F}_1$, $\mathcal{F}_2$ and a Bernoulli random variable $x$ which takes value $1$ w.p.\ $p$, $x\mathcal{F}_1\oplus(1-x)\mathcal{F}_2$ is a random variable such that it follows distribution $\mathcal{F}_1$ w.p.\ $p$, and follows $\mathcal{F}_2$  w.p.\ $1-p$. We use $\mathcal{N}(m,\sigma^2)$ to represent a Gaussian distribution with mean $m$ and variance $\sigma^2$.

A (finite) Markov Decision Process (MDP) is defined as a 6-tuple $\langle T,\gamma,S,A, \mathbf{p},\mathbf{r}\rangle$. Here, $T$ is the possibly infinite decision horizon; $\gamma\in(0,1]$ is the discount factor; $S$ is the  state set and $A_s$ is the  action set of state $s\in S$, both assumed to be finite. The parameter $\mathbf{p}$ and  $\mathbf{r}$ are the transition probability and the expected reward, respectively. That is, for $s\in S$ and $a\in A_s, r(s,a)$ is the expected reward and $p(s'|s,a)$ is the probability that the next state is $s'$. Following  \cite{puterman2009markov}, we denote the set of all history-dependent randomized strategies by $\Pi^{HR}$, and the set of all Markovian randomized strategies by $\Pi^{MR}$. We use subscript $s$ to denote the value associated with the state $s$, e.g., $\mathbf{r}_s$ denotes the vector form of the rewards associated with the state $s$, and $\pi_s$ is the (randomized) action chosen at state $s$ for strategy $\pi$. The elements in the vector $\mathbf{p}_s$ are listed in the following way: the transition probabilities of the same action are arranged in the same block, and inside each block they are listed according to the order of the next state. We use $\underline{s}$ to denote the (random) state following $s$, and $\Delta(s)$ to denote the probability simplex on $A_s$. We use $\bigotimes$ to represent Cartesian product, e.g., $\mathbf{p}=\bigotimes_{s\in S}\textbf{p}_s$.

For a given strategy $\pi\in \Pi^{HR}$, we denote the expected (discounted) total-reward under parameters pair $(\mathbf{p}, \mathbf{r})$ as 
$$u(\pi,\mathbf{p},\mathbf{r})\triangleq \mathbb{E}^{(\mathbf{p},\mathbf{r})}_{\pi}\left\{\sum_{i=1}^{T}\gamma^{i-1}r(s_i,a_i)\right\}.$$

A Distributionally Ambiguous MDP (DAMDP) is defined as a tuple $\langle T, \gamma, S, A, \tilde{\C}_S\rangle$, where the transition probability $ \mathbf{p}$ and the expected reward  $\mathbf{r}$  are unknown. Instead, they  are assumed to obey a joint distribution $\mu_0$ (also  unknown) that belongs to a known ambiguity set $\C_S\triangleq\prod_{(\mathbf{p},\mathbf{r})}\tilde{\C}_S$, where $\prod_{(\mathbf{p},\mathbf{r})}$ means taking the marginal distribution of $(\mathbf{p},\mathbf{r})$.  

While the DAMDP framework can be very general,  most $\tilde{\C}_S$ results in formulations computationally intractable (e.g., \cite{delage2010distributionally,xu2012distributionally}). Hence, our first requirement of $\tilde{\C}_S$ is that the parameters among different states are independent.
That is,
$$\tilde{\C}_S\triangleq\{\mu|\mu=\bigotimes_{s\in S}\mu_{s},\mu_s\in \tilde{\C}_s,\forall s\in S\},$$
where ``state-wise ambiguity set'' $\tilde{\C}_s$ is a set of distributions of parameters of state $s$. By the definition of $ \C_S$, the state-wise property applies to $\C_S$ as well. This property is same as the concept of ``s-rectangularity'' in \cite{wiesemann2013robust}, and is essential for reducing DAMDP to robust MDP in Lemma \ref{lemma1}. In addition, \cite{mannor2012ldst} showed that the robust MDP  with coupled uncertainty sets is computationally challenging, which implies solving DAMDP with nonrectangular ambiguity sets is even harder.


We now discuss the admissible state-wise ambiguity set. Our formulation of the state-wise ambiguity set follows the unifying framework of~\cite{wiesemann2013distributionally}. In specific, given $s\in S$, the state-wise ambiguity set is representable with the following standard form
\begin{equation}\label{C_s}
\tilde{\C}_s\triangleq\left\{\mu_s\left(\begin{aligned}
&\mathbf{p}_s\\&\mathbf{r}_s\\&\tilde{\mathbf{u}}_s\end{aligned}\right)\bigg|\begin{aligned}
&\mathbb{E}_{(\mathbf{p}_s,\mathbf{r}_s,\tilde{\mathbf{u}}_s)\sim\mu_s}[F_s \mathbf{p}_s+G_s\mathbf{r}_s+H_s\tilde{\mathbf{u}}_s]=\mathbf{c}_s,\\
&\mu_s\left(O_s^i\right)\in [\underline{\alpha}_s^i,\overline{\alpha}_s^i],\quad \forall i\in I_s\end{aligned}\right\}.
\end{equation}
Here, $F_s\in \mathbb{R}^{k\times(|A_s|\times|s|)},G_s\in \mathbb{R}^{k\times |A_s|}, H_s\in\mathbb{R}^{k\times Q}, \mathbf{c}_s\in \mathbb{R}^k$; $I_s=\{1,2,\dots, n_s\}$ is an index set and $O^i_s\subseteq \mathbb{R}^{|A_s|\times|s|}\times\mathbb{R}^{|A_s|}\times\mathbb{R}^Q$ is a set of possible values of the parameters $(\mathbf{p}_s,\mathbf{r}_s,\tilde{\mathbf{u}}_s)$, termed ``confidence set''; and $\underline{\alpha}_s^i, \overline{\alpha}_s^i\in [0,1]$, $\underline{\alpha}_s^i\leq\overline{\alpha}_s^i$ for all $i\in I_s$, is the lower bound and upper bound of the probability that the parameters belong to the confidence set. Thus, each confidence set $O^i_s$ provides an estimation of the uncertain parameters pair $(\mathbf{p}_s, \mathbf{r}_s,\tilde{\mathbf{u}}_s)$ subject to a different confidence level. Ambiguity sets $\tilde{\C}_s$ contain prescribed conic representable confidence sets and mean values residing on an affine manifold, which turns out to be rich enough to encompass and extend several ambiguity sets considered in the recent literature. We abuse the notation here to use $\tilde{\C}_s$ to denote the set of joint distributions of $(\mathbf{p}_s, \mathbf{r}_s,\tilde{\mathbf{u}}_s)$, and hence the set of joint distribution of $(\mathbf{p}_s, \mathbf{r}_s)$ is $\C_s\triangleq\prod_{(\mathbf{p}_s,\mathbf{r}_s)}\tilde{\C}_s$.

Notice that here we use a classical technique called ``lifting'':
we introduce an auxiliary random vector $\tilde{\mathbf{u}}$, so that some non-linear relationship can be modeled linearly. For example, as we see below, a constraint on the variance can be modeled using this standard form, which is otherwise impossible if auxiliary variable is not introduced. This lifting technique thus  allows us to model a rich variety of structural information about the marginal distribution of $(\mathbf{p},\mathbf{r})$ in a unified manner. 

Note that, when $n_s=1$ and the distribution set involving solo probability bound is of the form $\tilde{\C}_s=\{\mu_s(\mathbf{p}_s, \mathbf{r}_s,\tilde{\mathbf{u}}_s)|\mu_s\left(O^1_s\right)=1, \forall i\in I_s,s\in S\}$ (i.e., the distribution set only contains the support of random variables), the DAMDP reduces to the classical robust MDP formulation (\cite{iyengar2005robust,nilim2005robust}), where the a-priori information of unknown parameters is that they belong to an uncertainty set.

Assumption \ref{assumption2} to \ref{assumption1} are standard requirements for the confidence sets, proposed in \cite{wiesemann2013distributionally}. The first one asserts the relationship between different confidence sets.

\begin{assumption}[Nesting condition]\label{assumption2}
For any $s\in S$, all $i,i'\in I_s$ and $i\neq i'$, we have either $O^i_s \Subset O^{i'}_s$, $O^{i'}_s \Subset O^i_s$ or $O^i_s \cap O^{i'}_s=\emptyset$.
\end{assumption}

\begin{figure}[!t]
        \centering
        \subfloat[]{\includegraphics[width=1.8in]{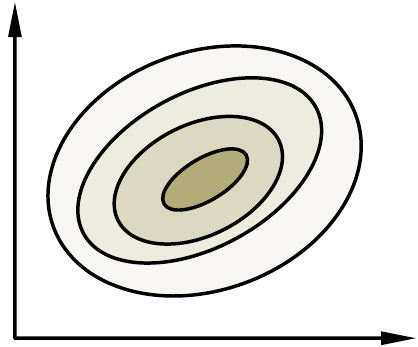}
        \label{figure1(a)}}
        \hfil
        \subfloat[]{\includegraphics[width=1.8in]{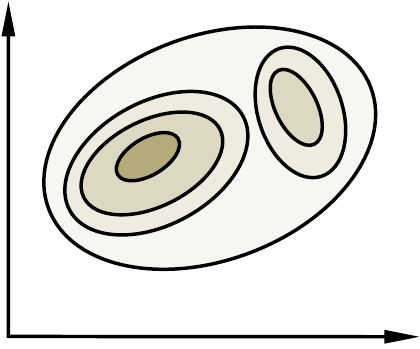}
        \label{figure1(b)}}
        \caption{Illustration of the confidence sets.}\label{figure1}
\end{figure}

Here $O^i_s \Subset O^{i'}_s$ means that a set $O^i_s$ is strictly included in a set $O^{i'}_s$, i.e., $O^i_s$ is contained in the interior of $O^{i'}_s$. The nesting condition is illustrated in Fig. \ref{figure1(b)}. The nesting condition implies a strict partial order on the confidence sets $O^i_s$ with respect to the $\Subset$-relation, with additional requirement that incomparable sets must be disjoint. Nested condition is a fundamental assumption needed for  the tractability of the distributionally robust optimization problems (cf.~\cite{wiesemann2013distributionally}). We remark here that when $\tilde{\C}_s$ is of form \eqref{C_s}, it trivially satisfies the nesting condition. Nesting condition can be verified analytically. Interested readers may refer to~\cite{wiesemann2013distributionally} Section 3.

We highlight the contributions of the paper by comparing with \cite{xu2012distributionally} who also propose a distributionally robust MDP approach. In~\cite{xu2012distributionally}, the state-wise ambiguity set is restricted to the following form:
$$\tilde{\C}_s=\{\mu_s|\mu_s\left(O_s^i\right)\geq \underline{\alpha}_s^i \quad \forall i=1,\dots, n_s\};$$
where $\underline{\alpha}_s^i\leq \underline{\alpha}_s^j, \,\, O_s^i\subseteq O_s^j,\,\, \forall i<j.$

Observe that our formulation  is more general as we allow additional linear constraints (in terms of expectation). This allows us to apply more statistical methods to estimate the uncertain parameters in MDPs. Moreover, their formulation requires a more restrictive nesting set condition where the confidence sets must have an incremental structure $O^1_s\subseteq O^2_s\subseteq \dots \subseteq O^{n_s}_s$, which is illustrated in Fig. \ref{figure1(a)}. While, as shown in Fig. \ref{figure1(b)}, the confidence sets we use can be disjoint and thus provide significantly more powerful modeling ability. This new structure is obviously more flexible, and can model more general distributions. For example, it can better characterize multimodal distributions such as mixed Gaussian distributions: We may establish confidence sets centered around corresponding peaks of different Gaussian components, as long as they do not intersect with each other. 

Next, for any $s\in S$  we  require that  $\tilde{\C}_s$ satisfy the following regularity condition.

\begin{assumption}[Regularity conditions for $\tilde{\C}_s$]\label{assumption4}
\begin{enumerate}[\IEEEsetlabelwidth{12)}]
\item The confidence set $O_s^{n_s}$ is bounded and has probability one, that is, $\underline{\alpha}_s^{n_s}=\overline{\alpha}_s^{n_s}=1$.
\item There is a probability distribution $\mu_s(\mathbf{p}_s, \mathbf{r}_s,\tilde{\mathbf{u}}_s)\in \tilde{\C}_s$ such that $\mu_s\left(O_s^i\right)\in (\underline{\alpha}_s^i,\overline{\alpha}_s^i)$ whenever $\underline{\alpha}_s^i<\overline{\alpha}_s^i$, $i\in I_s$.
\end{enumerate}
\end{assumption}

The condition 1 of Assumption \ref{assumption4} ensures that the confidence set with largest index, $O_s^{n_s}$, contains the support of the joint unknown parameters pair $(\mathbf{p}_s,\mathbf{r}_s,\tilde{\mathbf{u}}_s)$. The second condition stipulates that there is a probability distribution $\mu_s(\mathbf{p}_s, \mathbf{r}_s,\tilde{\mathbf{u}}_s)\in \tilde{\C}_s$ that satisfies the probability bounds in \eqref{C_s} as strict inequalities whenever the corresponding probability interval $[\underline{\alpha}_s^i,\overline{\alpha}_s^i]$ is non-degenerate. This assumption allows the strong duality to hold for {\em distributionally robust counterpart}, which we will define later. 

For each individual $O^i_s$, we make the following assumption for  tractability.

\begin{assumption}\label{assumption1}
For any $s\in S$ and $i\in I_s$, $O^i_s$ is nonempty and convex. Each confidence set $O^i_s$ is defined as 
\begin{equation*}
O^i_s=\left\{\left(\begin{aligned}
&\mathbf{p}_s\\&\mathbf{r}_s\\&\tilde{\mathbf{u}}_s\end{aligned}\right)\in\begin{aligned}
&\mathbb{R}^{|A_s|\times|s|}\times\mathbb{R}^{|A_s|}\times\mathbb{R}^Q\end{aligned}\bigg|\begin{aligned}&B_s^i\mathbf{p}_s+D_s^i\mathbf{r}_s+E_s^i\tilde{\mathbf{u}}_s\preceq_{K_s^{i}}\mathbf{b}_s^i\end{aligned}\right\},
\end{equation*}
where $B_s^i\in \mathbb{R}^{L_i\times (|A_s|\times|s|)}$, $D_s^i\in \mathbb{R}^{L_i\times |A_s|}$, $E_s^i\in\mathbb{R}^{L_i\times Q}$, $\mathbf{b}_s^i\in \mathbb{R}^{L_i}$, $K_s^i$ are proper cones (i.e., a closed, convex and pointed cone with nonempty interior).
\end{assumption}

Similarly to \cite{nilim2005robust}, we define the non-stationary model as: when a state is visited multiple times, each time it can take a different parameter realization. We define the stationary model as: uncertain parameters are chosen by nature depending on the decision maker's strategy once for all, and remain fixed thereafter. Each model leads to a game between the decision maker and nature, where the decision maker seeks to maximize the minimum expected reward, with the nature being the minimizing player. The second model is attractive for statistical reasons. Unfortunately, it turns out to be hard to solve. Therefore in this paper, for finite horizon case, we assume it shares the same property of non-stationary model: multiple visits to a state can be treated as visiting different states. This can easily be done by introducing dummy states. 

For finite horizon case, as~\cite{xu2012distributionally}, we make the following assumption, which will simplify our derivations.

\begin{assumption}\label{assumption3}
\begin{enumerate}[\IEEEsetlabelwidth{12)}]
\item Each state belongs to only one stage.
\item The terminal reward equals zero.
\item The first stage only contains one state $s^{ini}$.
\end{enumerate}
\end{assumption}

Using the condition 1 of Assumption \ref{assumption3}, we partition $S$ according to the stage each state belongs to. That is, we let $S_t$ be the set of states belong to $t^{th}$ stage. 

\section{Finite horizon distributionally robust MDPs}\label{sec3}
This section focuses on uncertain MDPs with a finite number of decision stages. We show that a strategy defined through backward induction, which we called S-robust strategy, is a distributionally robust strategy. We further show that such strategy is solvable in polynomial time under mild technical conditions. This thus generalizes similar results in~\cite{xu2012distributionally} to a significant more general class of ambiguity sets. 

For $\pi\in \Pi^{HR}$ and $\mu\in \C_S$, we denote the expected performance of a DAMDP as 
\begin{equation*}
\begin{aligned}
w(\pi,\mu,(s^{ini}))&\triangleq \mathbb{E}_{(\mathbf{p}, \mathbf{r})\sim\mu}\{u(\pi,\mathbf{p},\mathbf{r})\}=\int u(\pi,\mathbf{p},\mathbf{r})d\mu(\mathbf{p},\mathbf{r}).
\end{aligned}
\end{equation*}
\begin{definition}\label{definition1}
A strategy $\pi^*\in\Pi^{HR}$ is distributionally robust with respect to $\C_S$ if it satisfies that for all $\pi\in\Pi^{HR}$,
$$\inf_{\mu\in \C_S}w(\pi,\mu,(s^{ini}))\leq\inf_{\mu'\in \C_S}w(\pi^*,\mu',(s^{ini})).$$
\end{definition}

In words, each strategy is evaluated by its expected performance under the (respective) most adversarial distribution of the uncertain parameters, and a distributionally robust strategy is the optimal strategy according to this measure. 

The main focus of this section is deriving approaches to solve the distributionally robust strategy. To this end, we need the following definition.

\begin{definition}\label{definition2}
Given a DAMDP $\langle T, \gamma, S, A, \tilde{\C}_S\rangle$ with $T<\infty$, we define the S-robust problem as following
\begin{enumerate}
\item For $s\in S_T$, the S-robust value $\tilde{v}_T(s)\triangleq 0$.
\item For $s\in S_t$, where $t<T$, the S-robust value $\tilde{v}_t(s)$ and S-robust action $\tilde{\pi}_s$ are defined as
	\begin{equation}\label{S-robust problem}
		\begin{aligned}
			\tilde{v}_t(s)&\triangleq\displaystyle\
			\max_{\pi_s\in\Delta(s)}\{\min_{\mu_s\in \C_s}\mathbb{E}_{(\mathbf{p}_s, \mathbf{r}_s)\sim\mu_s}\{\mathbb{E}^{(\mathbf{p}_s,\mathbf{r}_s)}_{\pi_s}[r(s,a)+\gamma\tilde{v}_{t+1}(\underline{s})]\}\},\\
			\tilde{\pi}_s&\in
			\arg\displaystyle\max_{\pi_s\in\Delta(s)}\{\min_{\mu_s\in \C_s}\mathbb{E}_{(\mathbf{p}_s, \mathbf{r}_s)\sim\mu_s}\{\mathbb{E}^{(\mathbf{p}_s, \mathbf{r}_s)}_{\pi_s}[r(s,a)+\gamma\tilde{v}_{t+1}(\underline{s})]\}\}.
		\end{aligned}
	\end{equation}
			
\item A strategy $\tilde{\pi}^*$ is a S-robust strategy if $\forall s\in S$, and every history $h$ ends at $s$, we have $\tilde{\pi}_s^*$, conditioned on history $h$, is a S-robust action.
\end{enumerate}
\end{definition}
Note that the definition essentially requires that the strategy must be robust with respect to each sub-problem, and hence the name ``S-robust''. The following theorem shows any S-robust strategy $\pi^*$ is distributionally robust, and is the main result of this paper.

\begin{theorem}\label{theorem1}
Let $T<\infty$. Under Assumptions \ref{assumption2}, \ref{assumption1} and \ref{assumption3}, if $\pi^*$ is a S-robust strategy, then
\begin{enumerate}
\item $\pi^*$ is distributionally robust strategy with respect to $\C_S$.
\item There exists $\mu^*\in \C_s$ such that $(\pi^*,\mu^*)$ is a saddle point. That is,
\begin{equation*}
\begin{aligned}
\sup_{\pi\in \Pi^{HR}}w(\pi,\mu^*,(s^{ini}))&=w(\pi^*,\mu^*,(s^{ini}))=\inf_{\mu\in \C_S}w(\pi^*,\mu,(s^{ini})).
\end{aligned}
\end{equation*}
\end{enumerate}
\end{theorem}

\begin{IEEEproof}
The proof follows a similar structure as that of Theorem 3.1 in \cite{xu2012distributionally}: We first state a lemma from \cite{xu2012distributionally}, which shows that for a given strategy, the expected performance under admissible $\mu$ depends only on the expected value of the parameters. Thus we are able to reduce the distributionally robust MDPs to the classical robust MDPs. Then we show that the set of expected value of the uncertain parameters is convex and compact. Finally, by applying the results of classical robust MDPs we prove the theorem.

The following Lemma is indeed Lemma 3.2 of \cite{xu2012distributionally}. Hence we state the result without the proof. Interested readers may refer to \cite{xu2012distributionally} for the proof.
 
\begin{lemma}\label{lemma1}
Under the state-wise property of the ambiguity set $\tilde{\mathcal{C}}_S$, fix $\pi\in \Pi^{HR}$ and $\mu\in \C_S$, denote $\overline{\mathbf{p}}=\mathbb{E}_{\mu}(\mathbf{p})$ and  $\overline{\mathbf{r}}=\mathbb{E}_{\mu}(\mathbf{r})$. We have $$w(\pi,\mu,(s^{ini}))=u(\pi,\overline{\mathbf{p}},\overline{\mathbf{r}}).$$ 
\end{lemma}

Lemma \ref{lemma1} essentially means that for any strategy, the expected performance under an admissible distribution $\mu$ only depends on the expected value of the parameters under $\mu$. Thus, the distributionally robust MDPs reduces to the robust MDPs. Next we characterize the set of expected value of the parameters.

\begin{lemma}\label{lemma2}
For $s\in S$ and $\pi_s\in \Delta(s)$, we define the set 
$$\mathcal{Z}_s=\{\mathbb{E}_{\mu_s}(\mathbf{p}_s,\mathbf{r}_s)|\mu_s\in \C_s\}.$$ Then set $\mathcal{Z}_s$ is convex and compact.
\end{lemma}

\begin{IEEEproof} 
First, we show that, for $s\in S$ and $\pi_s\in \Delta(s)$, the set defined as
$\tilde{\mathcal{Z}}_s=\{\mathbb{E}_{\mu_s}(\mathbf{p}_s,\mathbf{r}_s,\tilde{\mathbf{u}}_s)|\mu_s\in \tilde{\C}_s\}$ is convex and compact.

Fix $s\in S$, and two distributions $\mu_s^1,\mu_s^2 \in \tilde{\C}_s$, and $\lambda\in[0,1]$, we have
\begin{equation*}
\begin{aligned}
&\lambda\mathbb{E}_{\mu_s^1}(\mathbf{p}_s, \mathbf{r}_s,\tilde{\mathbf{u}}_s)+(1-\lambda)\mathbb{E}_{\mu_s^2}(\mathbf{p}_s, \mathbf{r}_s,\tilde{\mathbf{u}}_s)=\mathbb{E}_{\lambda \mu_s^1 +(1-\lambda)\mu_s^2}(\mathbf{p}_s, \mathbf{r}_s,\tilde{\mathbf{u}}_s),
\end{aligned}
\end{equation*}
which holds due to the linearity of the expectation operation. Next, we show $\lambda\mu_s^1+(1-\lambda)\mu_s^2\in \tilde{\C}_s$. 
Since $\mu_s^1(\mathbf{p}_s^1,\mathbf{r}_s^1,\tilde{\mathbf{u}}_s^1),\mu_s^2(\mathbf{p}_s^2,\mathbf{r}_s^2,\tilde{\mathbf{u}}_s^2)\in \tilde{\C}_s$, for $s\in S$, we have,
\begin{equation*}
\begin{aligned}
&\lambda\int_{O_s^n}(F_s\mathbf{p}_s^1+G_s\mathbf{r}_s^1+H_s\tilde{\mathbf{u}}_s^1)d\mu_s^1(\mathbf{p}_s^1,\mathbf{r}_s^1,\tilde{\mathbf{u}}_s^1)+(1-\lambda)\int_{O_s^n}(F_s\mathbf{p}_s^2+G_s\mathbf{r}_s^2+H_s\tilde{\mathbf{u}}_s^2)d\mu_s^2(\mathbf{p}_s^2,\mathbf{r}_s^2,\tilde{\mathbf{u}}_s^2)
\\&=\lambda\mathbf{c}_s+(1-\lambda)\mathbf{c}_s=\mathbf{c}_s,\\
&\lambda\int_{O_s^n}\mathbf{1}_{[(\mathbf{p}_s^1,\mathbf{r}_s^1,\tilde{\mathbf{u}}_s^1)\in O^i_s]}d\mu_s^1(\mathbf{p}_s^1,\mathbf{r}_s^1,\tilde{\mathbf{u}}_s^1)+(1-\lambda)\int_{O_s^n}\mathbf{1}_{[(\mathbf{p}_s^2,\mathbf{r}_s^2,\tilde{\mathbf{u}}_s^2)\in O^i_s]}d\mu_s^2(\mathbf{p}_s^2,\mathbf{r}_s^2,\tilde{\mathbf{u}}_s^2)\\
&\leq\lambda\overline{\alpha}_s^i+(1-\lambda)\overline{\alpha}_s^i=\overline{\alpha}_s^i,\quad\forall i\in I_s,\\
&\lambda\int_{O_s^n}\mathbf{1}_{[(\mathbf{p}_s^1,\mathbf{r}_s^1,\tilde{\mathbf{u}}_s^1)\in O^i_s]}d\mu_s^1(\mathbf{p}_s^1,\mathbf{r}_s^1,\tilde{\mathbf{u}}_s^1)+(1-\lambda)\int_{O_s^n}\mathbf{1}_{[(\mathbf{p}_s^2,\mathbf{r}_s^2,\tilde{\mathbf{u}}_s^2)\in O^i_s]}d\mu_s^2(\mathbf{p}_s^2,\mathbf{r}_s^2,\tilde{\mathbf{u}}_s^2)\\
&\geq\lambda\underline{\alpha}_s^i+(1-\lambda)\underline{\alpha}_s^i=\underline{\alpha}_s^i,\quad\forall i\in I_s.
\end{aligned}
\end{equation*} 
Hence the convexity follows. To show that compactness, notice that $\tilde{\C}_s$ is weakly closed (i.e., closed w.r.t. to the weak topology) since the feasible set of each of constraint is weakly closed and so does their intersection. Thus, $\tilde{\mathcal{Z}}_s$ is closed since it is the image of $\tilde{\C}_s$ under expectation (which is a continuous function). This implies $\tilde{\mathcal{Z}}_s$ is compact since $O_s^{n_s}$ is bounded and hence $\tilde{\mathcal{Z}}_s$ is bounded.

Finally, since $\mathcal{Z}_s$ is the projection onto the first two coordinates of set $\tilde{\mathcal{Z}}_s$, its convexity and compactness are straightforward at this stage.
\end{IEEEproof}

Lemma \ref{lemma2} implies that, for $s\in S$ and $\pi_s\in \Delta(s)$, there exists $(\mathbf{p}_s^*, \mathbf{r}_s^*)\in \mathcal{Z}_s$ that satisfies 
$$\displaystyle\inf_{(\mathbf{p}_s, \mathbf{r}_s)\in\mathcal{Z}_s}u(\pi_s,\mathbf{p}_s,\mathbf{r}_s)=u(\pi_s,\mathbf{p}_s^*,\mathbf{r}_s^*).$$ 
Furthermore, we can construct $\bigotimes_{s\in S}\mathcal{Z}_s=\{\mathbb{E}_{\mu}(\mathbf{p},\mathbf{r})|\mu\in \C_S\}$ by state-wise decomposability of $\C_S$.  

We complete the proof of Theorem \ref{theorem1} by using the equivalence of distributionally robust MDPs and robust MDPs where the uncertainty set is $\bigotimes_{s\in S}\mathcal{Z}_s$. It is well known that for robust MDPs , saddle point of the minimax objective exists (\cite{nilim2005robust,iyengar2005robust}). More precisely, there exist 
$\pi^*\in\Pi^{HR},(\mathbf{p}^*, \mathbf{r}^*)\in \bigotimes_{s\in S}\mathcal{Z}_s$, such that 
\begin{equation*}
\begin{aligned}
\displaystyle\inf_{(\mathbf{p}, \mathbf{r})\in\bigotimes_{s\in S}\mathcal{Z}_s}u(\pi^*,\mathbf{p},\mathbf{r})&=u(\pi^*,\mathbf{p}^*,\mathbf{r}^*)=\sup_{\pi\in \Pi^{HR}}u(\pi,\mathbf{p}^*,\mathbf{r}^*)
\end{aligned}
\end{equation*}
holds. Moreover, we can construct $\pi^*$ and $(\mathbf{p}^*,\mathbf{r}^*)$ state-wise as $\pi^*=\bigotimes_{s\in S}\pi^*_s$ and $(\mathbf{p}^*,\mathbf{r}^*)=\bigotimes_{s\in S}(\mathbf{p}^*_s,\mathbf{r}^*_s)$. For each $s\in S_t$, $\pi^*_s$ and $(\mathbf{p}^*_s,\mathbf{r}^*_s)$ solves the zero-sum game
\begin{equation*}
\displaystyle\max_{\pi_s\in\Delta(s)}\min_{(\mathbf{p}_s, \mathbf{r}_s)\in \mathcal{Z}_s}\mathbb{E}^{(\mathbf{p}_s, \mathbf{r}_s)}_{\pi_s}[r(s,a)+\gamma\tilde{v}_{t+1}(\underline{s})].
\end{equation*}
Thus $\pi^*_s$ is any S-robust action, and hence $\pi^*$ can be any S-robust strategy. From Lemma \ref{lemma2}, there exists $\mu_s^*(\mathbf{p}^*_s, \mathbf{r}^*_s)\in \C_s$ satisfying $\mathbb{E}_{\mu^*_s}(\mathbf{p}_s, \mathbf{r}_s)=(\mathbf{p}^*_s, \mathbf{r}^*_s)$. Let $\mu^*=\bigotimes_{s\in S}\mu^*_s$. By applying Lemma \ref{lemma1}, we have 
\begin{equation*}
\begin{aligned}
&\sup_{\pi\in\Pi^{HR}}w(\pi,\mu^*,(s^{ini}))=\sup_{\pi\in \Pi^{HR}}u(\pi,\mathbf{p}^*,\mathbf{r}^*),\\
&w(\pi^*,\mu^*,(s^{ini}))=u(\pi^*,\mathbf{p}^*,\mathbf{r}^*),\\
&\inf_{\mu\in \C_S}w(\pi,\mu,(s^{ini}))=\displaystyle\inf_{(\mathbf{p}, \mathbf{r})\in\bigotimes_{s\in S}\mathcal{Z}_s}u(\pi^*,\mathbf{p},\mathbf{r}).
\end{aligned}
\end{equation*}
Above leads to
\begin{equation*}
\begin{aligned}
\sup_{\pi\in \Pi^{HR}}w(\pi,\mu^*,(s^{ini}))&=w(\pi^*,\mu^*,(s^{ini}))\\&=\inf_{\mu\in \C_S}w(\pi^*,\mu,(s^{ini})).
\end{aligned}
\end{equation*}
Thus, part $(2)$ of Theorem \ref{theorem1} holds. Part $(1)$ follows immediately.
\end{IEEEproof}

Define $\tilde{\mathbf{v}}_{t+1}$ as the vector form of $\tilde{v}_{t+1}(s')$ for all $s'\in S_{t+1}$, and 
$$\tilde{V}_s\triangleq 
	\begin{bmatrix}
       \tilde{\mathbf{v}}_{t+1}\mathbf{e}^\intercal_1(|A_s|)\\
       \vdots\\
       \tilde{\mathbf{v}}_{t+1}\mathbf{e}^\intercal_{|A_s|}(|A_s|) 
	\end{bmatrix}.$$	
Therefore, the expected reward under fixed parameter realization $(\mathbf{p}_s,\mathbf{r}_s)$ and any policy $\pi_s\in \Delta(s)$ is
	\begin{equation*}
	\begin{aligned}
		&\mathbb{E}^{(\mathbf{p}_s, \mathbf{r}_s)}_{\pi_s}[r(s,a)+\tilde{v}_{t+1}(\underline{s})]\\=&
		\sum_{a\in A_s}\pi_s(a)r(s,a)+\sum_{a\in A_s}\sum_{s'\in S_{t+1}}\pi_s(a)p(s'|s,a)\tilde{v}_{t+1}(s')\\
		=&\mathbf{r}_s^\intercal\pi_s+\mathbf{p}_s^\intercal\tilde{V}_s\pi_s.
	\end{aligned}
	\end{equation*}
	
We now investigate the computational aspect of the S-robust action.
			
\begin{theorem}\label{theorem2}
For $s\in S_t$ where $t<T$, the S-robust action is the optimal solution of the following optimization problem.
	\begin{equation*}
	\begin{aligned}
		\underset{w,\pi_s,\beta,\kappa,\lambda,\nu_i}{\text{minimize}}\quad & w \\
		\text{subject to}\quad & \mathbf{c}_s^\intercal\beta+\displaystyle\sum_{i\in I_s}[\overline{\alpha}_s^i\kappa_i-\underline{\alpha}_s^i\lambda_i]\leq w \\
		&\nu_i^\intercal \mathbf{b}_s^i-\sum_{i'\in \textit{A}(i)}[\kappa_{i'}-\lambda_{i'}]\leq 0,\quad i\in I_s\\						
		&{B_s^i}^\intercal\nu_i+\tilde{V}_s\pi_s+F_s^\intercal\beta=0,\quad i\in I_s\\
		&{D_s^i}^\intercal\nu_i+\pi_s+G_s^\intercal\beta=0,\quad i\in I_s\\
        &{E_s^i}^\intercal\nu_i+H_s^\intercal\beta=0,\quad i\in I_s\\
		&\pi_s\in \Delta(s),\\
		&\beta\in \mathbb{R}^k,\quad\kappa,\lambda\in \mathbb{R}^{n_s}_{+},\quad\nu_i\in {K_s^i}^*.
	\end{aligned}
	\end{equation*}
			
Here, ${K_s^i}^*$ represents the cone dual to $K_s^i$, and set $\textit{A}(i)\triangleq \{i\}\cup\{i'\in I_s:O^i_s \Subset O^{i'}_s\}$.
\end{theorem}

\begin{IEEEproof}
First, for $s\in S_t$, since the expectation term in \eqref{S-robust problem} is independent of the auxiliary random vector $\tilde{\mathbf{u}}_s$, the S-robust problem can be written as
	\begin{equation*}
	\begin{aligned}
		\underset{w,\pi_s}{\text{\textit{minimize}}}\quad & w \\
		\text{\textit{subject to}}\quad & \max_{\mu_s\in \tilde{\C}_s}-\mathbb{E}_{(\mathbf{p}_s, \mathbf{r}_s)\sim\mu_s}\{\mathbb{E}^{(\mathbf{p}_s, \mathbf{r}_s)}_{\pi_s}[r(s,a)+\gamma\tilde{v}_{t+1}(\underline{s})]\}\leq w\\
		&\pi_s\in \Delta(s).\\
	\end{aligned}
	\end{equation*}
			
Now we consider the constraint, which we call {\em distributionally robust counterpart}:
$$ \max_{\mu_s\in \tilde{\C}_s}-\mathbb{E}_{(\mathbf{p}_s, \mathbf{r}_s)\sim\mu_s}\{\mathbb{E}^{(\mathbf{p}_s, \mathbf{r}_s)}_{\pi_s}[r(s,a)+\gamma\tilde{v}_{t+1}(\underline{s})]\}\leq w.$$
The {\em distributionally robust counterpart} can be equivalently expressed as below
\begin{equation*}
 	\left.\begin{aligned}
        \underset{\mu_s}{\text{\textit{maximize}}}\quad & -\int_{O_s^{n_s}}\mathbb{E}^{(\mathbf{p}_s, \mathbf{r}_s)}_{\pi_s}[r(s,a)+\gamma\tilde{v}_{t+1}(\underline{s})]d\mu_s(\mathbf{p}_s,\mathbf{r}_s,\tilde{\mathbf{u}}_s)\\
       	\text{\textit{subject to}}\quad & \int_{O_s^{n_s}} (F_s \mathbf{p}_s+G_s\mathbf{r}_s +H_s\tilde{\mathbf{u}}_s)d\mu_s(\mathbf{p}_s,\mathbf{r}_s,\tilde{\mathbf{u}}_s)=\mathbf{c}_s\\
       	&\int_{O_s^{n_s}}\mathbf{1}_{[(\mathbf{p}_s,\mathbf{r}_s,\tilde{\mathbf{u}}_s)\in O^i_s]}d\mu_s(\mathbf{p}_s,\mathbf{r}_s,\tilde{\mathbf{u}}_s)\geq \underline{\alpha}_s^i,\forall i\in I_s\\
		&\int_{O_s^{n_s}}\mathbf{1}_{[(\mathbf{p}_s,\mathbf{r}_s,\tilde{\mathbf{u}}_s)\in O^i_s]}d\mu_s(\mathbf{p}_s,\mathbf{r}_s,\tilde{\mathbf{u}}_s)\leq \overline{\alpha}_s^i,\forall i\in I_s
       \end{aligned}
 	\right\}
	 \quad\leq w.
\end{equation*}
			
Theorem 2 of~\cite{wiesemann2013distributionally}  states that for the above maximization problem, the strong duality holds under Assumption \ref{assumption2} and \ref{assumption4}. The dual problem is of the following form:

	\begin{equation}\label{semi-infinite problem}
 	\left.\begin{aligned}
		\underset{\beta,\kappa,\lambda}{\text{\textit{minimize}}}\quad & \mathbf{c}_s^\intercal\beta+\displaystyle\sum_{i\in I_s}[\overline{\alpha}_s^i \kappa_i-\underline{\alpha}_s^i\lambda_i]\\
		\text{\textit{subject to}}\quad &[F_s \mathbf{p}_s+G_s\mathbf{r}_s+H_s\tilde{\mathbf{u}}_s]^\intercal\beta+\displaystyle\sum_{i'\in \textit{A}(i)}[\kappa_{i'}-\lambda_{i'}]\geq
		-\mathbb{E}^{(\mathbf{p}_s,\mathbf{r}_s)}_{\pi_s}[r(s,a)+\gamma\tilde{v}_{t+1}(\underline{s})],\\&\qquad\qquad\qquad\qquad\qquad\qquad\qquad\qquad\qquad\qquad\forall(\mathbf{p}_s, \mathbf{r}_s,\tilde{\mathbf{u}}_s)\in O^i_s,\forall i\in I_s\\
		&\beta\in \mathbb{R}^k,\quad\kappa,\lambda\in \mathbb{R}^{n_s}_{+}
       \end{aligned}
 	\right\}
	 \quad\leq w,
	\end{equation}
where $\textit{A}(i)\triangleq \{i\}\cup\{i'\in I_s:O^i_s \Subset O^{i'}_s\}$.

The $i^{th}$ semi-infinite constraint in the above problem can be written as, for each $i\in I_s$, 
	\begin{equation*}
	 \left.\begin{aligned}
			\underset{\mathbf{p}_s,\mathbf{r}_s,\tilde{\mathbf{u}}_s}{\text{\textit{maximize}}}\quad & -\{[F_s \mathbf{p}_s+G_s\mathbf{r}_s+H_s\tilde{\mathbf{u}}_s]^\intercal\beta+\sum_{i'\in\textit{A}(i)}[\kappa_{i'}-\lambda_{i'}]+\mathbf{r}_s^\intercal\pi_s+\mathbf{p}_s^\intercal\tilde{V}_s\pi_s\}\\
			\text{\textit{subject to}}\quad & B_s^i\mathbf{p}_s+D_s^i\mathbf{r}_s+E_s^i\tilde{\mathbf{u}}_s\preceq_{K_s^{i}}\mathbf{b}_s^i.\\
       \end{aligned}
	 \right\}
	 \quad\leq 0.
	\end{equation*}

The above problem is a  convex optimization problem. By applying the duality theory of convex optimization~\cite{boyd2004convex}, we get its dual below. For each $i\in I_s$,
	\begin{equation*}
	 \left.\begin{aligned}
			\underset{\nu_i}{\text{\textit{minimize}}}\quad & \nu_i^\intercal \mathbf{b}_s^i-\sum_{i'\in \textit{A}(i)}[\kappa_{i'}-\lambda_{i'}]\\
			\text{\textit{subject to}}\quad 			&{B_s^i}^\intercal\nu_i+\tilde{V}_s\pi_s+F_s^\intercal\beta=0\\
			&{D_s^i}^\intercal\nu_i+\pi_s+G_s^\intercal\beta=0\\
			&{E_s^i}^\intercal\nu_i+H_s^\intercal\beta=0\\
			&\nu_i\in {K_s^i}^*.
       \end{aligned}
 	\right\}
 	\quad\leq 0.
	\end{equation*}
Thus, the $i^{th}$ semi-infinite constraint can be equivalently reformulated as follows: there exists $\nu_i\in{K_s^i}^*$,  $C_i^\intercal\nu_i+\tilde{V}_s\pi_s+A^\intercal\beta=0$, $D_i^\intercal\nu_i+\pi_s+B^\intercal\beta=0$ and ${E_s^i}^\intercal\nu_i+H_s^\intercal\beta=0$ such that $\nu_i^\intercal \mathbf{b}_s^i-\sum_{i'\in \textit{A}(i)}[\kappa_{i'}-\lambda_{i'}]\leq 0$. Finally, by combining constraints for all $i\in I_s$, we obtain the optimization formulation stated in the theorem.  
\end{IEEEproof}

Thus, since for $s\in S_t$, $\Delta(s)$ is compact, we can solve the S-robust action in polynomial time if all $K^i_s$ are ``easy'' cones such as linear, conic quadratic  or semidefinite cones. Moreover, using Theorem~\ref{theorem1}, by backward induction, we can obtain the S-robust strategy efficiently. 

By virtue of the lifting technique \cite[Thm 5]{wiesemann2013distributionally}, we show below that several widely used ambiguity sets are indeed special cases of $\tilde{\C}_s$ defined in \eqref{C_s}. We further list their corresponding S-robust problems.

\begin{example}[Mean]
Assume that we only know a noisy empirical estimator of the exact mean of $\mathbf{p}_s$. That is, given $G\in \mathbb{R}^{M\times (|A_s|\times |s|)}$, $\mathbf{f} \in \mathbb{R}^{M}$ and $\mathbf{p}_s\sim \mu_s(\mathbf{p}_s)$, $G\mathbb{E}_{\mathbf{p}_s\sim\mu_s(\mathbf{p}_s)}$ $[\mathbf{p}_s]\preceq_{K}\mathbf{f}$, where $K$ is a proper cone. \cite{wiesemann2013distributionally} shows that $\tilde{\C}_s$, which involves the auxiliary random vector $\tilde{\mathbf{u}}_s\in \mathbb{R}^{M}$, can be expressed as
$$\tilde{\C}_s=\left\{\mu_s(\mathbf{p}_s,\tilde{\mathbf{u}}_s)\bigg|\begin{aligned}&\mathbb{E}_{\tilde{\mathbf{u}}_s\sim\mu_s}[\tilde{\mathbf{u}}_s]=\mathbf{f},\\&\mu_s\left(G\mathbf{p}_s\preceq_{K}\tilde{\mathbf{u}}_s\right)=1\end{aligned}\right\}.$$ 

Note that $\mu_s(\mathbf{p}_s)\in\prod_{\mathbf{p}_s}\tilde{\C}_s$, where $\prod_{\mathbf{p}_s}\tilde{\C}_s$ denotes the marginal distribution of $\mathbf{p}_s$ under the joint probability distribution of two random vectors $\mathbf{p}_s$ and $\tilde{\mathbf{u}}_s$. The problem in Theorem \ref{theorem2} now takes the form
	\begin{equation*}
	\begin{aligned}
		\underset{w,\pi_s,\kappa,\nu}{\text{minimize}}\quad & w \\
		\text{subject to}\quad & \kappa+\mathbf{f}^\intercal\nu\leq w \\
		&\kappa+\mathbf{r}_s^\intercal\pi_s\geq0\\
		&\tilde{V}_s\pi_s+G^\intercal\nu=0\\
		&\pi_s\in\Delta(s),\quad\nu\in K^*.
	\end{aligned}
	\end{equation*}	
The same argument can be developed for mean of $\mathbf{r}_s$. This example can also be treated via ``classical'' robust optimization models by virtue of Lemma \ref{lemma1}.
\end{example}

\begin{example}[Variance]
This example imposes conditions on both mean and covariance of the distribution. First, we assume that the mean of the random reward $\mathbf{r}_s\sim \mu_s(\mathbf{r}_s)$ is given by $\mathbb{E}_{\mathbf{r}_s\sim\mu_s(\mathbf{r}_s)}[\mathbf{r}_s]=\mathbf{m}$, and $\mathbb{E}_{\mathbf{r}_s\sim\mu_s(\mathbf{r}_s)}[(\mathbf{r}_s-\mathbf{m})(\mathbf{r}_s-\mathbf{m})^{\intercal}]\preceq\varSigma$ for $\varSigma\in\mathbb{S}^{|A_s|}_+$. We denote the ambiguity set under fixed mean $\mathbf{m}$ as $\tilde{\C}_s(\mathbf{m})$. As discussed in \cite{wiesemann2013distributionally}, $\tilde{\C}_s(\mathbf{m})$, which involves a random matrix $\tilde{U}_s\in \mathbb{R}^{|A_s|\times|A_s|}$, can be expressed as
$$\tilde{\C}_s(\mathbf{m})=\left\{\mu_s\left(\begin{aligned}&\mathbf{r}_s\\&\tilde{U}_s\end{aligned}\right)\bigg|\begin{aligned}
&\mathbb{E}_{\mathbf{r}_s\sim\mu_s}[\mathbf{r}_s]=\mathbf{m},\quad\mathbb{E}_{\tilde{U}_s\sim\mu_s}[\tilde{U}_s]=\varSigma,\\
&\mu_s\left(\begin{bmatrix}
 1 & (\mathbf{r}_s-\mathbf{m})^{\intercal} \\
	          (\mathbf{r}_s-\mathbf{m}) & \tilde{U}_s \\   
     \end{bmatrix}\succeq0\right)=1\end{aligned}\right\}.$$
We have $\mu_s(\mathbf{r}_s)\in\prod_{\mathbf{r}_s}\tilde{\C}_s(\mathbf{m})$, where $\prod_{\mathbf{r}_s}\tilde{\C}_s(\mathbf{m})$ denotes the marginal distribution of $\mathbf{r}_s$ under the joint probability distribution of random variables $\mathbf{r}_s$ and $\tilde{U}_s$. Next, we consider the case where given $G\in \mathbb{R}^{M\times|A_s|}$, $\mathbf{f} \in \mathbb{R}^{M}$, the mean of random reward $\mathbf{r}_s$ is restricted by $G\mathbf{m}\leq\mathbf{f}$. In this case, the ambiguity set can be written as $\tilde{\C}_s=\bigcup_{G\mathbf{m}\leq\mathbf{f}}\tilde{\C}_s(\mathbf{m})$. By Theorem \ref{theorem2}, the S-robust action can be solved using the following formulation
	\begin{equation*}
	\begin{aligned}
		\underset{w,\pi_s,\kappa,\nu,\gamma,Y}{\text{minimize}}\quad & w \\
		\text{subject to}\quad & \kappa+\textbf{tr}(\Sigma Y_{22}^{\intercal})-\mathbf{f}^{\intercal}\nu\leq w \\
		&\kappa+\mathbf{p}_s^\intercal\tilde{V}_s\pi_s-Y_{11}-\mathbf{f}^{\intercal}\gamma\geq0\\
		&\pi_s-Y_{12}-Y_{21}-G^{\intercal}\nu=0\\
		&Y_{12}+Y_{21}+G^{\intercal}\gamma=0\\
		&Y-\begin{bmatrix}
		       Y_{11} & Y_{12}^{\intercal} \\
		       Y_{21} & Y_{22} \\
		     \end{bmatrix}=0\\
		&\pi_s\in\Delta(s),\quad\nu,\gamma\geq0,\quad Y\in\mathbb{S}_+^{|A_s|+1}.\\
	\end{aligned}
	\end{equation*}
	
The same argument can be developed for variance of $\mathbf{p}_s$. 
\end{example}

\begin{example}[Mean Absolute Deviation]
Assume that $\mathbb{E}_{\mathbf{r}_s\sim\mu_s(\mathbf{r}_s)}$ $[|\mathbf{r}_s-
\mathbf{m}|] \leq\mathbf{f}$ for $\mathbf{m},\mathbf{f}\in\mathbb{R}^{|A_s|}$. \cite{wiesemann2013distributionally} shows that $\tilde{\C}_s$, which involves the auxiliary random vector $\tilde{\mathbf{u}}_s\in \mathbb{R}^{|A_s|}$, can be expressed as
$$\tilde{\C}_s=\left\{\mu_s(\mathbf{r}_s,\tilde{u}_s)\bigg|\begin{aligned}&\mathbb{E}_{\tilde{\mathbf{u}}_s\sim\mu_s}[\tilde{\mathbf{u}}_s]=\mathbf{f},\\&\mu_s
\left(\begin{aligned}&\tilde{\mathbf{u}}_s\geq\mathbf{r}_s-\mathbf{m}\\&\tilde{\mathbf{u}}_s\geq\mathbf{m}-\mathbf{r}_s\end{aligned}\right)=1\end{aligned}\right\}.$$ 
     
Note that $\mu_s(\mathbf{r}_s)\in\prod_{\mathbf{r}_s}\tilde{\C}_s$, where $\prod_{\mathbf{r}_s}\tilde{\C}_s$ denotes the marginal distribution of $\mathbf{r}_s$ under the joint probability distribution of random variables $\mathbf{r}_s$ and $\tilde{\mathbf{u}}_s$. The problem in Theorem \ref{theorem2} in this case can be represented as
	\begin{equation*}
	\begin{aligned}
		\underset{w,\pi_s,\kappa,\nu}{\text{minimize}}\quad & w \\
		\text{subject to}\quad & \kappa-\mathbf{f}^\intercal\nu\leq w \\
		&\kappa+\mathbf{p}_s^\intercal\tilde{V}_s\pi_s+\mathbf{m}^\intercal\pi_s\geq0\\
		&\pi_s\in \Delta(s),\quad\nu\geq0.\\
	\end{aligned}
	\end{equation*}
	
The corresponding S-robust problem can be obtained for mean absolute deviation of $\mathbf{p}_s$. 
\end{example}

\begin{example}[Expected Huber Loss Function]
For a scalar $z\in\mathbb{R}$ the Huber Loss function is defined as
\[H_{\delta}(z) = \left\{ 
  \begin{array}{l l}
    \frac{1}{2}z^2 & \quad \text{if $|z|\leq\delta$},\\
    \delta(|z|-\frac{1}{2}\delta) & \quad \text{otherwise,}
  \end{array} \right.\]
where $\delta>0$ is a prescribed robustness parameter.

Assume that $\mathbb{E}_{\mathbf{r}_s\sim\mu_s(\mathbf{r}_s)}[H_{\delta}(\mathbf{f}^{\intercal}\mathbf{r}_s)]\leq g$ for $\mathbf{f}\in\mathbb{R}^{|A_s|}$ and $g\in\mathbb{R}_{+}$. \cite{wiesemann2013distributionally} shows that $\tilde{\C}_s$, which involves the auxiliary random variables $\tilde{u}_s,\tilde{v}_s,\tilde{w}_s,\tilde{s}_s,\tilde{t}_s\in\mathbb{R}_{+}$, can be expressed as

$$\tilde{\C}_s=\left\{\mu_s\left(\begin{aligned}
&\mathbf{r}_s,\tilde{u}_s,\\&\tilde{v}_s,\tilde{w}_s,\\&\tilde{s}_s,\tilde{t}_s\end{aligned}\right)\bigg|\begin{aligned}&\mathbb{E}_{\tilde{w}_s\sim\mu_s}[\tilde{w}_s]=g,\\&\mu_s\left(\begin{aligned}&\delta(\tilde{u}_s-\tilde{s}_s)+\frac{\tilde{s}_s^2}{2}+\delta(\tilde{v}_s-\tilde{t}_s)+\frac{\tilde{t}_s^2}{2}\leq\tilde{w}_s,\\&\tilde{u}_s\geq\tilde{s}_s,\quad\tilde{v}_s\geq\tilde{t}_s,\quad\mathbf{f}^{\intercal}\mathbf{r}_s=\tilde{u}_s-\tilde{v}_s
\end{aligned}\right)=1\end{aligned}\right\}.$$ 

Note that $\mu_s(\mathbf{r}_s)\in\prod_{\mathbf{r}_s}\tilde{\C}_s$, where $\prod_{\mathbf{r}_s}\tilde{\C}_s$ denotes the marginal distribution of $\mathbf{r}_s$ under the joint probability distribution of six random variables. The S-robust problem in Theorem \ref{theorem2} is
	\begin{equation*}
	\begin{aligned}
		\underset{w,\pi_s,\kappa,\nu,\gamma,\eta,\theta}{\text{minimize}}\quad & w \\
		\text{subject to}\quad & 
		\kappa-g\nu\leq w \\
		&\kappa+\mathbf{p}_s^\intercal\tilde{V}_s\pi_s+3\delta\gamma+3\delta\eta-3\delta^2\nu-\frac{3\gamma^2}{2\nu}-\frac{3\eta^2}{2\nu}\geq0\\
        &\pi_s+\theta\mathbf{f}=0\\
		&\gamma-\delta\nu-\theta=0\\
		&\eta-\delta\nu+\theta=0\\
		&\pi_s\in\Delta(s),\quad\nu,\gamma,\eta\geq0.
	\end{aligned}
	\end{equation*}
	
The corresponding S-robust problem for the case of transition probability can be obtained in a similar way. 
\end{example}

Before concluding this section, we briefly discuss the computational complexity. We denote $M$ as the maximal computational effort in calculating possibly randomized S-robust action $\pi_s$ for each $s\in S_t$. Thus, the total time-complexity is $\mathcal{O}(T|S|M)$.

\begin{remark}[Stationary \& non-stationary model]
The S-robust value we derived in this section is for non-stationary model. It provides a lower bound for stationary model, which is generally hard to solve. Thus, one can use the non-stationary model to approximate the stationary model, when the latter is intractable.
\end{remark}

\section{Infinite horizon distributionally robust MDPs}\label{sec4}

In this section, we study the distributionally ambiguous MDP in the discounted-reward infinite horizon setup. In specific, we 
generalize the notion of S-robust strategy proposed in Section 3 to discounted-reward infinite horizon case, and show that the S-robust strategy is distributionally robust. Similarly to \cite{xu2012distributionally}, we consider two models, namely, the non-stationary model and the stationary model. The non-stationary model treats the system as having infinitely many states, each visited at most once. Therefore, we consider an equivalent MDP with an augmented state space, where each augmented state is defined by a pair $(s,t)$ where $s\in S$ and $t$, meaning state $s$ in the $t^{th}$ horizon. We define the set of distributions as the Cartesian product of the admissible distribution of each (augmented) state. That is,
$$\overline{\C}^{\infty}_S\triangleq \left\{\mu|\mu=\bigotimes_{s\in S,t=1,2,\dots}\mu_{s,t};\mu_{s,t}\in \C_s,\forall s\in S,\quad\forall t=1,2,\dots\right\}.$$
The stationary model treats the system as having a finite number of states, while multiple visits to one state is allowed. That is, if a state $s$ is visited for multiple times, then each time the distribution (of uncertain parameters) $\mu_s$ is the same. We define the set of admissible distribution as
$$\overline{\C}_S\triangleq \left\{\mu|\mu=\bigotimes_{s\in S,t=1,2,\dots}\mu_{s,t};\mu_{s,t}=\mu_{s},\mu_{s}\in \C_s,\forall s\in S,\quad\forall t=1,2,\dots\right\}.$$
These two formulations can model different setups: if the system, more specifically the distribution of uncertain parameters, evolves with time, then non-stationary model is more appropriate; while if the system is static, then stationary model is preferable.

The S-robust strategy for infinite horizon distributionally robust MDPs is defined as follows.
\begin{definition}\label{definition4}
Given a DAMDP  $\langle T, \gamma, S, A, \tilde{\C}_S\rangle$ with $T=\infty$, we define the S-robust problem as following
\begin{enumerate}
\item The S-robust value $\tilde{v}_{\infty}(s)$ and S-robust action $\tilde{\pi}_s$ are defined as
			\begin{equation*}
			\begin{aligned}
			\tilde{v}_{\infty}(s)&\triangleq  \displaystyle\max_{\pi_s\in\Delta(s)}\{\min_{\mu_s\in \C_s}\mathbb{E}_{(\mathbf{p}_s, \mathbf{r}_s)\sim\mu_s}\{\mathbb{E}^{(\mathbf{p}_s, \mathbf{r}_s)}_{\pi_s}[r(s,a)+\gamma\tilde{v}_{\infty}(\underline{s})]\}\},\\
			\tilde{\pi}_s&\in\arg\displaystyle\max_{\pi_s\in\Delta(s)}\{\min_{\mu_s\in \C_s}\mathbb{E}_{(\mathbf{p}_s, \mathbf{r}_s)\sim\mu_s}\{\mathbb{E}^{(\mathbf{p}_s,\mathbf{r}_s)}_{\pi_s}[r(s,a)+\gamma\tilde{v}_{\infty}(\underline{s})]\}\}.
			\end{aligned}
			\end{equation*}						
\item A strategy $\tilde{\pi}^*$ is a S-robust strategy if $\forall s\in S$, $\tilde{\pi}_s^*$ is a S-robust action.
\end{enumerate}
\end{definition}
To see that the S-robust strategy is well defined, it suffices to show that the following operator $\mathcal{L}:\mathbb{R}^{|S|}\rightarrow\mathbb{R}^{|S|}$ is a $\gamma$ contraction with respect to $\ell_{\infty}$ norm.
$$\{\mathcal{L}\mathbf{v}\}(s)\triangleq\max_{\pi_s\in\Delta(s)}\min_{\mu_s\in \C_s}\{\mathcal{L}^{\mu_s}_{\pi_s}\mathbf{v}\}(s),$$ 
where for $\mathbf{v}_1,\mathbf{v}_2$, given any $\pi_s\in \Delta(s)$, and $\mu_s\in \C_s$,
\begin{equation*}
\begin{aligned}
&\{\mathcal{L}^{\mu_s}_{\pi_s}\mathbf{v}\}(s)\triangleq\mathbb{E}_{(\mathbf{p}_s,\mathbf{r}_s)\sim\mu_s}\{\mathbb{E}^{(\mathbf{p}_s,\mathbf{r}_s)}_{\pi_s}[r(s,a)+\gamma v(\underline{s})]\\
&=\sum_{a\in A_s}\pi_s(a)r(s,a)+\gamma\sum_{a\in A_s}\sum_{s'\in S}\pi_s(a)p(s'|s,a)v(s').
\end{aligned}
\end{equation*}
\begin{lemma}\label{lemma3}
Under Assumption \ref{assumption1}, for $0\leq\gamma<1$, $\mathcal{L}$ is a $\gamma$ contraction with respect to $\ell_{\infty}$ norm.
\end{lemma}
\begin{IEEEproof}
For arbitrary $\mathbf{v}_1,\mathbf{v}_2$, and fix $s\in S$. Let $\pi_s^1,\mu_s^1(\mathbf{p}^1_s, \mathbf{r}_s^1)$ and $\pi_s^2,\mu_s^2(\mathbf{p}^2_s, \mathbf{r}_s^2)$ be the respective maximizing and minimizing variables. Assume $\{\mathcal{L}\mathbf{v}_1\}(s)\geq\{\mathcal{L}\mathbf{v}_2\}(s)$, we have
\begin{equation*}
\begin{array}{l l}
0&\leq\{\mathcal{L}\mathbf{v}_1\}(s)-\{\mathcal{L}\mathbf{v}_2\}(s)\\&=\{\mathcal{L}^{\mu_s^1}_{\pi_s^1}\mathbf{v}_1\}(s)-\{\mathcal{L}^{\mu_s^2}_{\pi_s^2}\mathbf{v}_2\}(s)\\
&\leq\{\mathcal{L}^{\mu_s^2}_{\pi_s^1}\mathbf{v}_1\}(s)-\{\mathcal{L}^{\mu_s^2}_{\pi_s^1}\mathbf{v}_2\}(s).
\end{array}
\end{equation*}

The right hand side term can be expressed as
\begin{equation*}
\begin{array}{l l}
&\{\mathcal{L}^{\mu_s^2}_{\pi_s^1}\mathbf{v}_1\}(s)-\{\mathcal{L}^{\mu_s^2}_{\pi_s^1}\mathbf{v}_2\}(s)\\&=\gamma\displaystyle\sum_{a\in A_s}\sum_{s'\in S}\pi_s^1(a)p_2(s'|s,a)(v_1(s')-v_2(s'))\\
&\leq\gamma\displaystyle\sum_{a\in A_s}\sum_{s'\in S}\pi_s^1(a)p_2(s'|s,a)\|\mathbf{v}_1-\mathbf{v}_2\|_\infty\\
&=\gamma\|\mathbf{v}_1-\mathbf{v}_2\|_\infty.
\end{array}
\end{equation*}

Last equality holds due to the fact that $\pi_s^1$ is a vector on the probability simplex and $\sum_{s'\in S}p_2(s'|s,a)=1$. Thus, for any $\mathbf{v}_1,\mathbf{v}_2$, given any $\pi_s\in \Delta(s)$, and $\mu_s\in \C_s$, $\mathcal{L}^{\mu_s}_{\pi_s}$ is a $\gamma$ contraction.

The above derivation establishes that when $\{\mathcal{L}\mathbf{v}_1\}(s)\geq\{\mathcal{L}\mathbf{v}_2\}(s)$ we have $0\leq\{\mathcal{L}\mathbf{v}_1\}(s)-\{\mathcal{L}\mathbf{v}_2\}(s)\leq\gamma\|\mathbf{v}_1-\mathbf{v}_2\|_\infty$. Repeating this argument in the case that $\{\mathcal{L}\mathbf{v}_1\}(s)\leq\{\mathcal{L}\mathbf{v}_2\}(s)$ implies that
$$|\{\mathcal{L}\mathbf{v}_1\}(s)-\{\mathcal{L}\mathbf{v}_2\}(s)|\leq\gamma\|\mathbf{v}_1-\mathbf{v}_2\|_\infty$$
for all $s\in S$. Taking the supreme over $s$ in the above expression yields the result.
\end{IEEEproof}

Note that for given any $\mathbf{v}$ and each $s$, by applying Theorem \ref{theorem2}, S-robust strategy can be obtained efficiently. Banach Fixed-Point Theorem states that, there exists a unique $\mathbf{v}^*$ such that $\mathcal{L}\mathbf{v}^*=\mathbf{v}^*$, which is the S-robust value by definition. Moreover, for arbitrary $\mathbf{v}^0$, the value vector sequence defined by $\mathcal{L}^n\mathbf{v}^0=\mathbf{v}^n$ converges exponentially to $\mathbf{v}^*$. Therefore, as the following lemma shows, we can compute the S-robust action for each $s$ (and hence S-robust strategy) using this procedure.

\begin{lemma}\label{lemma4}
Given $s\in S$. Let $\mathbf{v}^n=\mathcal{L}^n\mathbf{v}^0$, and 
$$\begin{aligned}\pi^n_s\in \arg\displaystyle\max_{\pi_s\in\Delta(s)}\{\min_{\mu_s\in \C_s}&\mathbb{E}_{(\mathbf{p}_s,\mathbf{r}_s)\sim\mu_s}\{\mathbb{E}^{(\mathbf{p}_s,\mathbf{r}_s)}_{\pi_s}[r(s,a)+\gamma v^n(\underline{s})]\}\}.\end{aligned}$$
Then the sequence $\{\pi_s^n\}_{n=1}^{\infty}$ has convergent subsequences, and any of its limiting points is a S-robust action of state $s$.
\end{lemma}
\begin{IEEEproof}
Since $\Delta(s)$ is compact, the sequence $\{\pi_s^n\}_{n=1}^{\infty}$ has convergent subsequences. To show that any limiting point is a S-robust action, without loss of generality, we assume $\pi_s^n\rightarrow\pi_s^*$. 

 
For $\mathbf{v}$, $\mathbf{v}'$, and given any $\mu_s\in \C_s$, $\hat{\pi}_s\in \Delta(s)$, we note that $\mathcal{L}^{\mu_s}_{\hat{\pi}_s}$ is a $\gamma$ contraction (see the proof of Lemma \ref{lemma3}), that is,
$$\{\mathcal{L}^{\mu_s}_{\hat{\pi}_s}\mathbf{v}\}(s)\leq\{\mathcal{L}^{\mu_s}_{\hat{\pi}_s}\mathbf{v}'\}(s)+\gamma\|\mathbf{v}-\mathbf{v}'\|_{\infty}.$$
By definition, for any $\hat{\pi}_s\in \Delta(s)$, we have
$$\min_{\mu_s\in \C_s}\{\mathcal{L}^{\mu_s}_{\hat{\pi}_s}\mathbf{v}^n\}(s)\leq\min_{\mu_s'\in \C_s}\{\mathcal{L}^{\mu_s'}_{\pi^n_s}\mathbf{v}^n\}(s).$$

Therefore, for $\mathbf{v}^*$ defined by $\mathcal{L}\mathbf{v}^*=\mathbf{v}^*$, we have
\begin{equation}\label{equ1}
\begin{array}{l l l}
&&\displaystyle\min_{\mu_s\in \C_s}\{\mathcal{L}^{\mu_s}_{\pi^n_s}\mathbf{v}^*\}(s)-\min_{\mu_s\in \C_s}\{\mathcal{L}^{\mu_s}_{\hat{\pi}_s}\mathbf{v}^*\}(s)\\
&=&\{\displaystyle\min_{\mu_s\in \C_s}\{\mathcal{L}^{\mu_s}_{\pi^n_s}\mathbf{v}^*\}(s)-\displaystyle\min_{\mu_s\in \C_s}\{\mathcal{L}^{\mu_s}_{\pi^n_s}\mathbf{v}^n\}(s)\}\\
&&+\{\displaystyle\min_{\mu_s\in \C_s}\{\mathcal{L}^{\mu_s}_{\pi^n_s}\mathbf{v}^n\}(s)-\displaystyle\min_{\mu_s\in \C_s}\{\mathcal{L}^{\mu_s}_{\hat{\pi}_s}\mathbf{v}^*\}(s)\}\\
&\geq&\{\displaystyle\min_{\mu_s\in \C_s}\{\mathcal{L}^{\mu_s}_{\pi^n_s}\mathbf{v}^*\}(s)-\displaystyle\min_{\mu_s\in \C_s}\{\mathcal{L}^{\mu_s}_{\pi^n_s}\mathbf{v}^n\}(s)\}\\
&&+\{\displaystyle\min_{\mu_s\in \C_s}\{\mathcal{L}^{\mu_s}_{\pi^n_s}\mathbf{v}^n\}(s)-\displaystyle\min_{\mu_s\in \C_s}\{\mathcal{L}^{\mu_s}_{\pi^n_s}\mathbf{v}^*\}(s)\}\\
&\geq&-2\gamma\|\mathbf{v}^n-\mathbf{v}^*\|_{\infty}.
\end{array}
\end{equation}

By Lemma \ref{lemma2}, we denote $\mu_s^*=\arg\min_{\mu_s\in \C_s}\{\mathcal{L}^{\mu_s}_{\hat{\pi}_s}\mathbf{v}^*(s)\}$. We have
\begin{equation}\label{equ2}
\begin{aligned}
\lim\limits_{n\rightarrow\infty}\min_{\mu_s\in \C_s}\{\mathcal{L}^{\mu_s}_{\pi^n_s}\mathbf{v}^*\}(s)&\leq\lim\limits_{n\rightarrow\infty}\{\mathcal{L}^{\mu_s^*}_{\pi^n_s}\mathbf{v}^*\}(s)\\&=\{\mathcal{L}^{\mu_s^*}_{\pi^*_s}\mathbf{v}^*\}(s)\\&=\min_{\mu'_s\in \C_s}\{\mathcal{L}^{\mu'_s}_{\pi^*_s}\mathbf{v}^*\}(s).
\end{aligned}
\end{equation}

Here the first equality holds since $\{\mathcal{L}^{\mu_s}_{\pi_s}\mathbf{v}\}(s)$ is continuous on $\pi_s$ and $\pi_s^n\rightarrow\pi_s^*$, and the second equality holds due to the definition of $\mu_s^*$. Combine \eqref{equ1} and \eqref{equ2}, and note that $\mathbf{v}^n\rightarrow\mathbf{v}^*$ due to Lemma \ref{lemma3}, we obtain
$$\min_{\mu_s'\in \C_s}\{\mathcal{L}^{\mu_s'}_{\pi^*_s}\mathbf{v}^*\}(s)\geq\min_{\mu_s\in \C_s}\{\mathcal{L}^{\mu_s}_{\hat{\pi}_s}\mathbf{v}^*\}(s).$$

Hence, $\pi^*_s$ is a S-robust action of state $s$.
\end{IEEEproof}

Based on these lemmas, we have the following two theorems which show that the S-robust strategy is distributionally robust. Indeed, these two theorems are similar to Theorem 4.1 and 4.2 of \cite{xu2012distributionally}. We omit the proofs as they are identical to those of \cite{xu2012distributionally}.

\begin{theorem}\label{theorem3}
Under Assumption \ref{assumption1}, given $T=\infty$ and $0\leq\gamma<1$, any S-robust strategy is distributionally robust with respect to $\overline{\C}_S^{\infty}$.
\end{theorem}

\begin{theorem}\label{theorem4}
Under Assumption \ref{assumption1}, given $T=\infty$ and $0\leq\gamma<1$, any S-robust strategy is distributionally robust with respect to $\overline{\C}_S$.
\end{theorem}

In terms of computation, to achieve an accuracy of $\varepsilon$, the computational complexity of infinite horizon case should be $\mathcal{O}(T|S|M\log\varepsilon/\log\gamma)$. 

\begin{remark}[Stationary \& non-stationary model]
For any given strategy, Theorem \ref{theorem3} and \ref{theorem4} implies that, when $T\rightarrow\infty$, the distributionally robust strategies for both formulations coincide, and can be computed by iteratively solving the S-robust problem defined in Definition \ref{definition4}.
\end{remark}

\section{Simulation}\label{sec5}
In this section, we study two synthetic numerical examples: a machine replacement problem and a path planning problem. In the machine replacement problem, the reward is assumed to be uncertain; whereas in the path planning problem, the transition probability is uncertain. Our numerical simulation shows  that by incorporating more general probabilistic information, the proposed distributionally robust approach handles uncertainty in a more flexible way, and hence leads to a better performance than the nominal, robust and distributionally robust approach proposed in \cite{xu2012distributionally}. All results were generated on an Intel Core i5-3570 CPU with 3.40 GHz clock speed and 8 GB RAM. The S-robust problems in robust and distributionally robust approaches are solved in Matlab using the CVX package \cite{cvx}.

\subsection{Reward uncertainty in the machine replacement problem}

We consider a machine replacement problem similar to that considered in \cite{delage2010percentile}. Consider the repair cost incurred by a factory that holds a high number of machines, given that each of these machines is modeled with a same underlying MDP for which rewards are subject to uncertainty. Note that we select two simple instances of machine replacement problem to better illustrate how our method compares with the nominal and the (distributionally) robust approaches. In fact, our method remains computationally tractable for much larger scale machine replacement problems with more than 1,000 states.

\subsubsection{Machine replacement as a MDP with Gaussian rewards}

\begin{figure}[!t]
	\centering
	\includegraphics[width=3.5in]{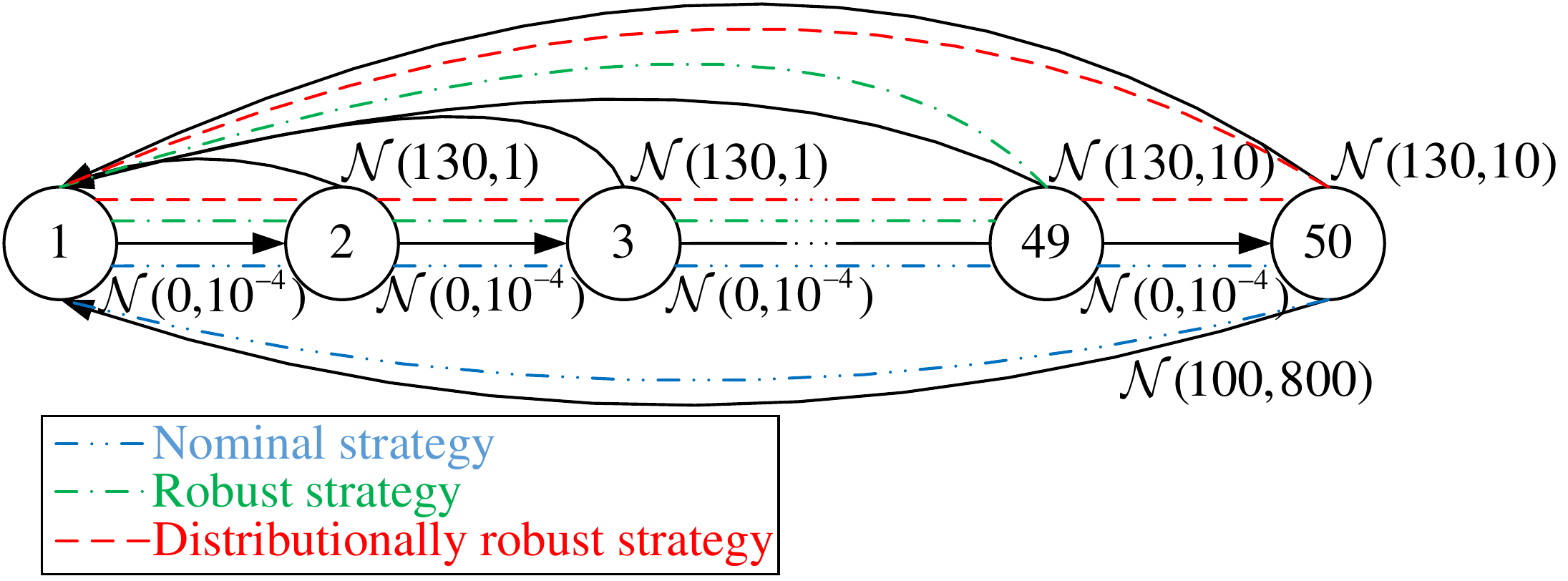}
	\caption{Instance of a machine replacement problem
	with Gaussian uncertainty in the rewards.}\label{figure2}
\end{figure}

In our first experiment with Gaussian rewards MDP, we consider a machine replacement problem with 50 states, 2 actions (``repair'' and ``not repair'') for each state, deterministic transitions, a discount factor of 0.8, and Gaussian distribution of the uncertain rewards (see Fig. \ref{figure2}). For each of the first 48 steps, the ``repair'' action has a cost independently distributed as $\mathcal{N}(130,1)$. The 49th and 50th states of the machine's life are designed to be risky: not repairing at state 50 incurs a highly uncertain cost $\mathcal{N}(100,800)$, while repairing at both states is a more secure but still uncertain option with a cost $\mathcal{N}(130,10)$. The detailed implementation is as follows:
By observing sample data generated from the true Gaussian distribution, we are able to compute the estimated mean $\hat{m}$ and variance $\hat{\sigma}^2$. We use the mean value of uncertain rewards to compute the nominal strategy. For both robust and distributionally robust strategy, we construct confidence sets using $\hat{m}\pm 3\hat{\sigma}$ for the first 49 states, and $\hat{m}\pm 4\hat{\sigma}$ for state 50, as it is more risky and thus hard to estimate. In addition, we construct an extra $62.32\%$ confidence interval $O_{50}^1$ (centered at the mean) with $60\%-70\%$ confidence level (i.e., $\underline{\alpha}_{50}^1=0.6$ and $\overline{\alpha}_{50}^1=0.7$) for distributionally robust strategy. The optimal paths followed for three strategies are shown in Fig. \ref{figure2}.

\begin{figure}[ht]
	\centering
	\includegraphics[width=5in]{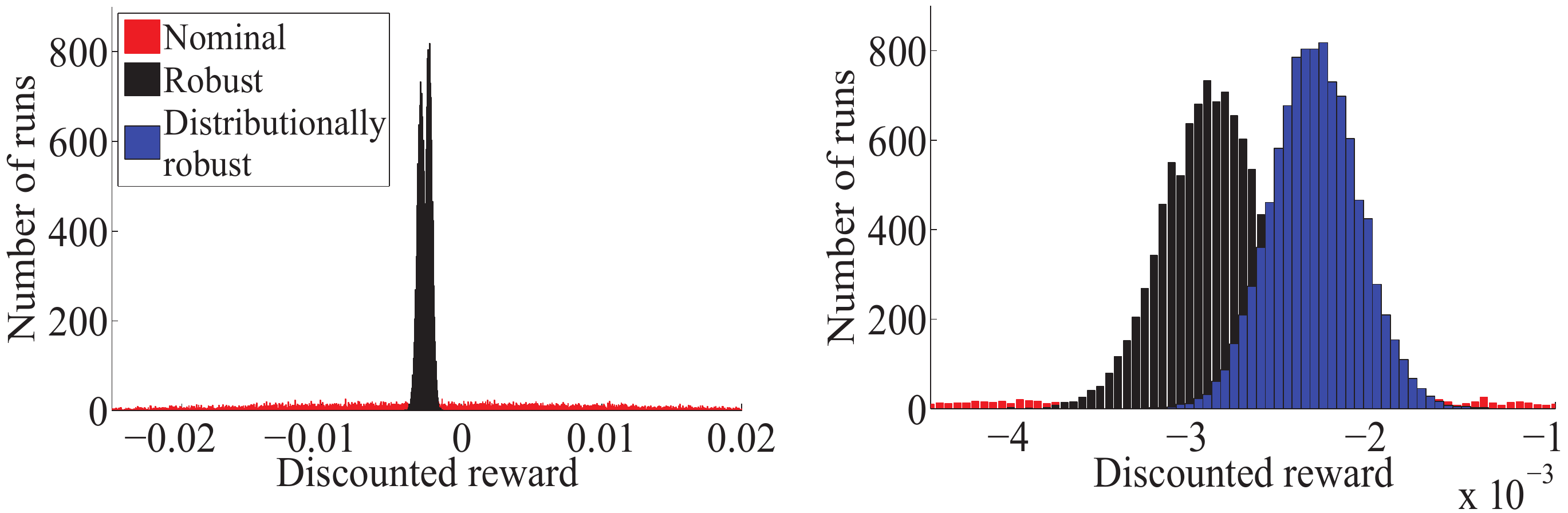}
	\caption{Performance comparisons between nominal, robust, and distributionally robust strategies on 10,000 runs of the machine replacement problem with Gaussian rewards (The right figure focuses on the interval $[-0.0045,-0.001]$).}\label{figure3}
\end{figure}

The performance of the strategies obtained by using the nominal, the robust and the distributionally robust approaches is presented in Fig. \ref{figure3}. The corresponding average total discounted rewards and computational times are shown in Table \ref{table1}. The nominal strategy results in the highest average total discounted rewards. This is well expected as we are using the exact mean value of the reward as the nominal parameter. However, the nominal strategy is highly risky: it cannot prevent the bad preformance (e.g., $-2.5\times10^{-2}$) from happening, which is undesirable. These results  coincide with what one would typically expect from the three solution concepts. While the nominal strategy, blind to any form of risk, finds no advantage in ever repairing, the robust strategy ends up following a highly conservative policy (repairing the machine at state 49 to avoid state 50). On the other hand, the distributionally robust optimal strategy makes use of more distributional information and handles the risk efficiently by waiting until state 50 and then repair the machine. Therefore, this strategy beats the nominal and robust strategies in that it strikes a good tradeoff between high mean reward and low variance over 10,000 different trials. These results coincide with what one would typically expect from the three solution concepts.

\begin{table}[!t]
\caption{Average total discounted rewards and computational times of nominal, robust, and distributionally robust strategies in machine replacement problem with Gaussian rewards}\label{table1}
\centering
\begin{tabular}{|c|c|c|c|}
\hline
\multirow{2}{*}{}                & \multicolumn{3}{c|}{Strategies}                                     \\ \cline{2-4} 
                                 & Nominal             & Robust              & Distributionally robust \\ \hline
Average total discounted rewards & $-1.8\times10^{-3}$ & $-2.9\times10^{-3}$ & $-2.3\times10^{-3}$     \\ \hline
Computational times (seconds)    & $0.643$             & $815$               & $820$                   \\ \hline
\end{tabular}
\end{table}

\subsubsection{Machine replacement as a MDP with mixed Gaussian rewards}

\begin{figure}[!t]
	\centering
	\includegraphics[width=3.5in]{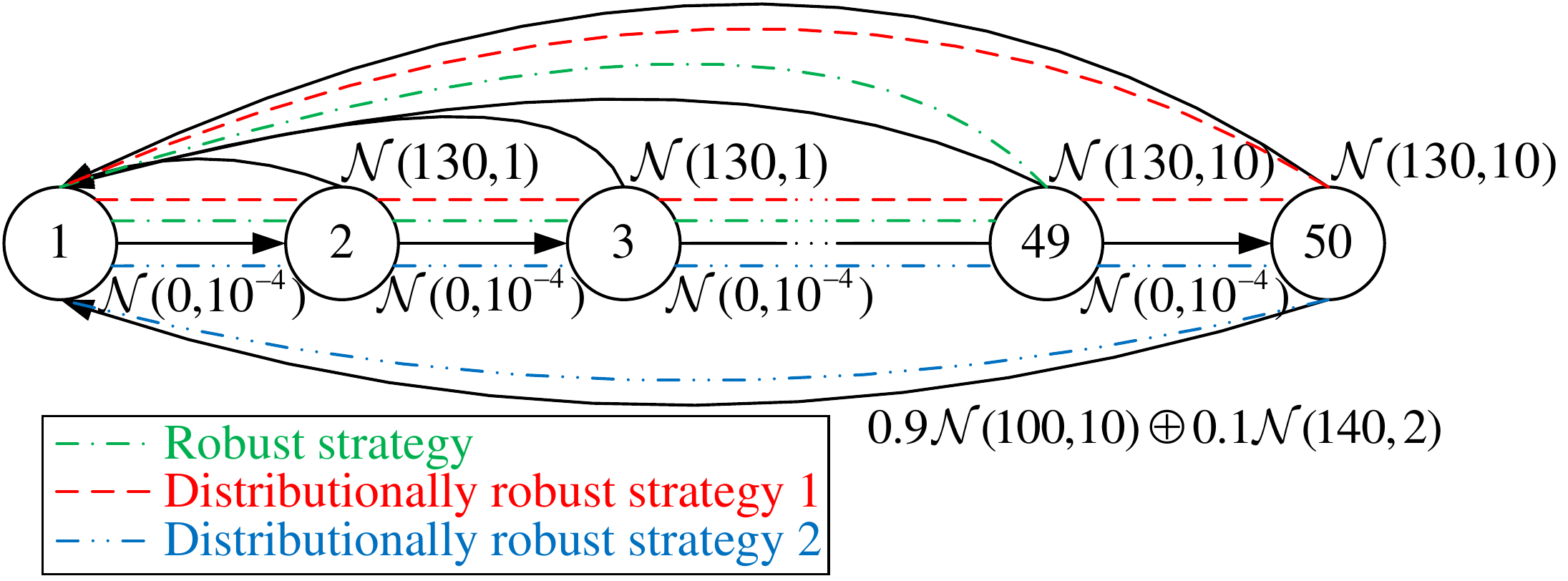}
	\caption{Instance of a machine replacement problem
		with mixed Gaussian uncertainty in the rewards.}\label{figure4}
\end{figure}

\begin{figure}[!t]
        \centering
        \subfloat[]{\includegraphics[width=2in]{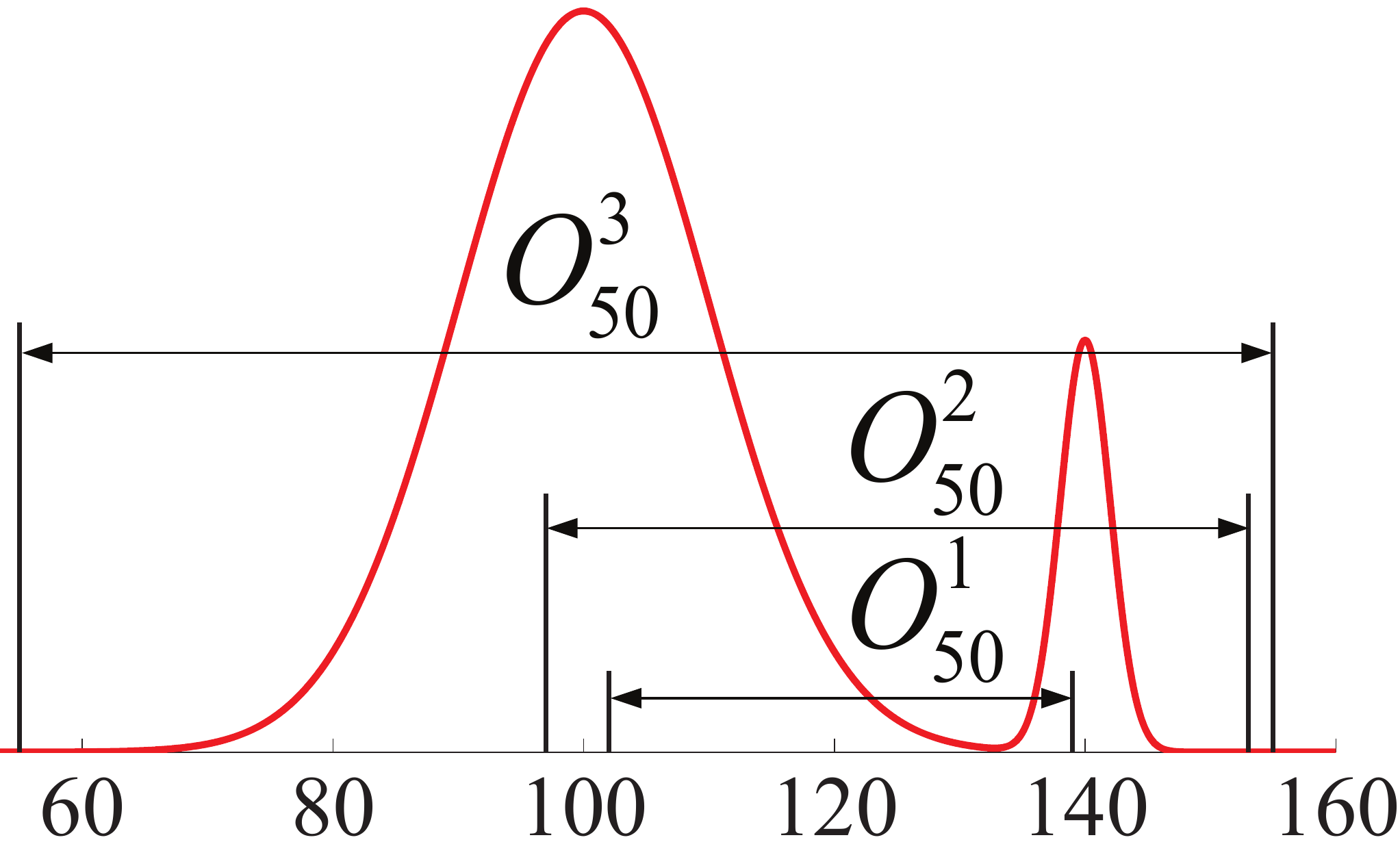}
	    \label{figure5(a)}}
	    \hfil
        \subfloat[]{\includegraphics[width=2in]{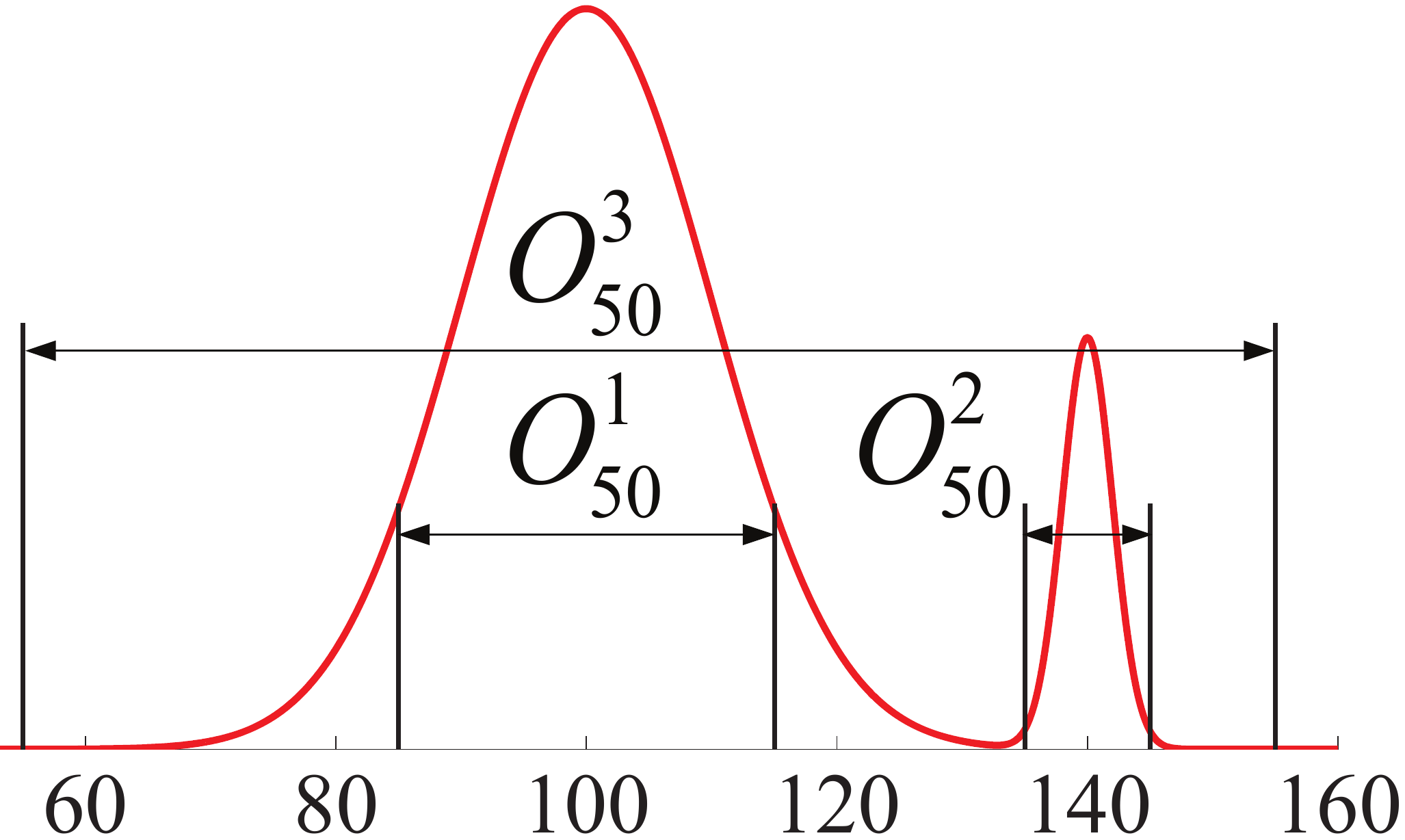}
        \label{figure5(b)}}
        \caption{Illustration of the confidence sets for two distributionally robust strategies.}\label{figure5}
\end{figure}

The second experiment considers a similar setup as the previous one, except that not repairing at the 50th state now  has a reward which follows a {\em mixed Gaussian distribution} (see Fig. \ref{figure4}). This experiment illustrates the effect of the two different nested-set structures shown in Fig. \ref{figure1}. In specific, we  apply the two different distributionally robust approaches (proposed in this paper and in \cite{xu2012distributionally} respectively) to this problem, and show that our method, which is more general, archives better performance. The detailed implementation is as follows: For the Gaussian mixture model, the estimated mean and variance for each Gaussian cannot be computed by simply taking mean and variance of the entire sample data set. The observed data must be divided into several Gaussians, each of with its own mean and variance. We obtain the parameters for each Gaussian by applying the expectation-maximization algorithm \cite{moon1996expectation}. Note that the algorithm requires an initial guess as to how many Gaussians are hidden in the distribution. This can be done by checking the histogram of the observed sample data. We choose it to be 2 in this experiment. After we get the estimated means and variances, the estimated probability density function can be obtained. For the robust and two distributionallly robust strategies, we construct uncertainty set corresponding to  $99\%$ probability support of the rewards for the first 49 states, and $99.9\%$ for the 50th state that is more risky, using the estimated probability density function. For the first distributionally robust strategy (the one proposed in \cite{xu2012distributionally}), we construct two additional {\em nested} confidence sets of the uncertain rewards (see Fig. \ref{figure5(a)}):  $40.96\%$ confidence set $O_{50}^1$ with $40\%-50\%$ confidence level (lower/upper confidence bounds $\underline{\alpha}_{50}^1=0.4$, $\overline{\alpha}_{50}^1=0.5$);  $65.61\%$ confidence set $O_{50}^2$ with $60\%-70\%$ confidence level (lower/upper confidence bounds $\underline{\alpha}_{50}^2=0.6$, $\overline{\alpha}_{50}^2=0.7$). In contrast, for the second distributionally robust strategy (the one proposed in this paper), we construct two {\em disjoint} confidence sets of the uncertain rewards (see Fig. \ref{figure5(b)}): $77.97\%$ confidence set $O_{50}^1$ with $70\%-80\%$ confidence level (lower/upper confidence bounds $\underline{\alpha}_{50}^1=0.7$, $\overline{\alpha}_{50}^1=0.8$), and $9.90\%$ confidence set $O_{50}^2$ with $0\%-10\%$ confidence level (lower/upper confidence bounds $\underline{\alpha}_{50}^2=0, \overline{\alpha}_{50}^2=0.1$). Specifically, we select these two intervals around the peaks of the two Gaussian elements (i.e., $\mathcal{N}(100,10)$ and $\mathcal{N}(140,2)$) to better model this mixed distribution. The optimal paths followed for three strategies are shown in Fig. \ref{figure4}. 

\begin{figure}[!t]
	\centering
	\includegraphics[width=2.8in]{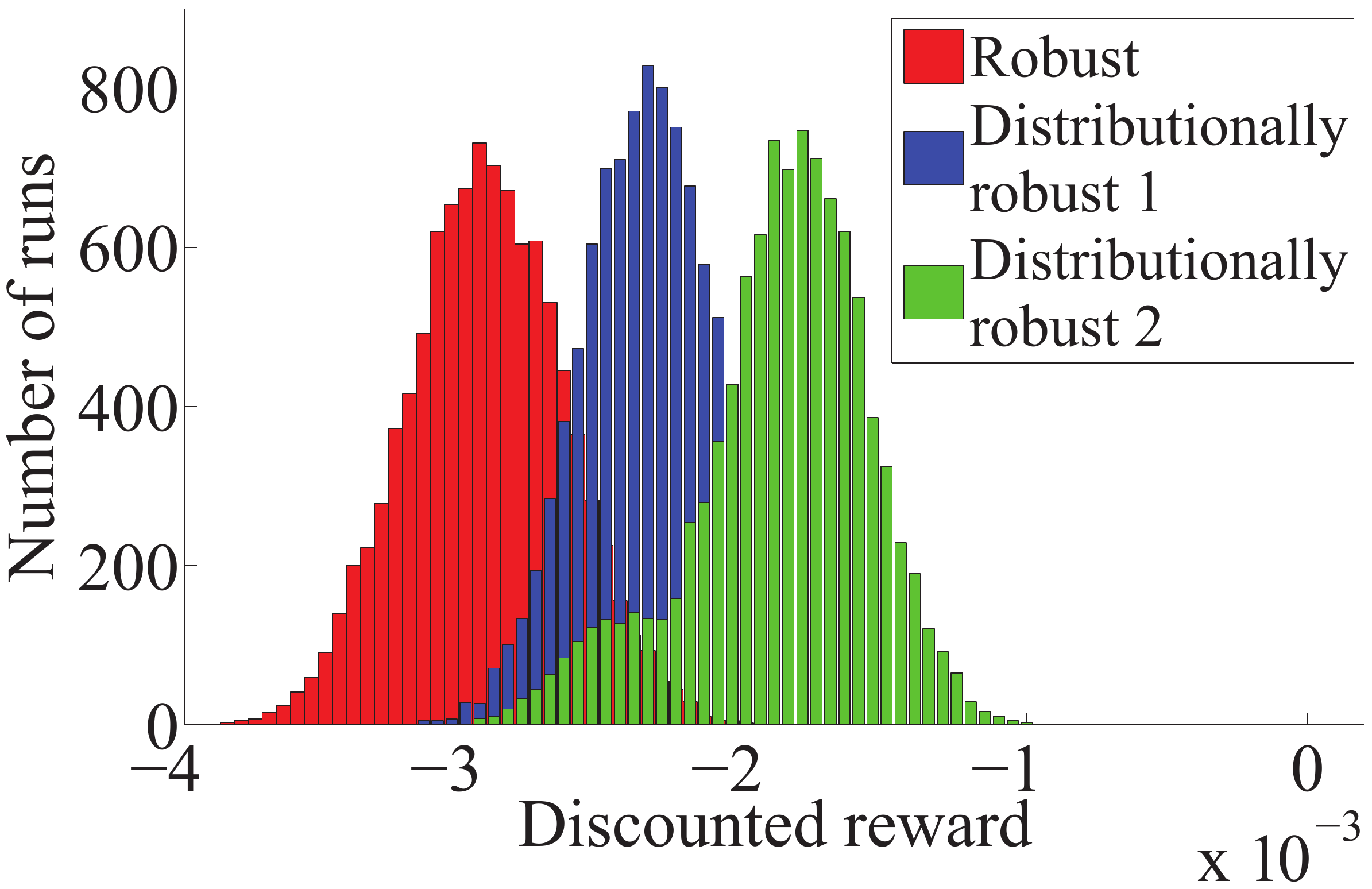}
	\caption{Performance comparisons between robust and two distributionally robust strategies on 10,000 runs of the machine replacement problem with mixed Gaussian rewards.}\label{figure6}
\end{figure}

The performance of the strategies obtained by using robust and two distributionally robust approaches is presented in Fig. \ref{figure6}. The corresponding average total discounted rewards and computational times are shown in Table \ref{table2}. As expected, the robust strategy ends up following a highly conservative policy which repairs the machine at state 49 to avoid state 50. The first distributionally robust strategy, not modeling the mixture Gaussian distribution well, finds it advantageous to repair at the 50th state. In contrast, capable of capturing the distribution information in a more flexible way, the second distributionally robust strategy better models the uncertainty and finds that not repairing the machine at state 50 is optimal. The performance comparison clearly shows that the second distributionally robust strategy is more desirable, which highlights that the distributionally robust approach with general structure of confidence sets can be beneficial in practice.

We remark that, in practice, one can obtain the modality structure of uncertain parameters in a data-driven way by applying clustering algorithms to an initial primitive data set. For example, one may check the histogram of historical observations. If the data concentrates on several distinct and disjoint bins, our multi-model DAMDP approach can be applied.

\begin{table}[h]
\caption{Average total discounted rewards and computational times of robust and two distributionally robust strategies in machine replacement problem with mixed Gaussian rewards}\label{table2}
\centering
\begin{tabular}{|c|c|c|c|}
\hline
\multirow{2}{*}{}                & \multicolumn{3}{c|}{Strategies}                                            \\ \cline{2-4} 
                                 & Robust              & Distributionally robust 1 & Distributionally robust 2 \\ \hline
Average total discounted rewards & $-2.9\times10^{-3}$ & $-2.3\times10^{-3}$       & $-1.9\times10^{-3}$      \\ \hline
Computational times (seconds)    & $849$               & $862$                     & $820$                    \\ \hline
\end{tabular}
\end{table}

\subsection{Transition uncertainty in the path planning problem}

\begin{figure}[!t]
	\centering
	\includegraphics[width=3.5in]{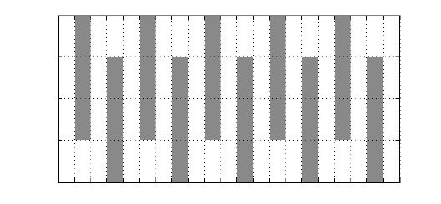}
	\caption{The maze for the path planning problem}\label{figure7}
\end{figure}

In this subsection, we consider a path planning problem, similar to the one presented in~\cite{xu2012distributionally}: an agent wants to exit a $4\times21$ maze (shown in Fig. \ref{figure7}) using the least possible time. Starting from the upper-left corner, the agent can move up, down, left and right, but can only exit the grid at the lower-right corner. Here, a white box stands for a normal place where the agent needs one time unit to pass through. A shaded box represents a ``shaky'' place: if an agent reaches a ``shaky'' place, then he may risk jumping to the starting point (``reboot''). The true transition probability of the jump follows a distribution $(1-\lambda)\mathcal{N}(0.1,10^{-4})\oplus\lambda\mathcal{N}(0.2,10^{-4})$ where $\lambda\in (0,1]$. The four approaches are implemented as follows: The nominal approach neglects this random jump. The robust approach takes a worst-case analysis, i.e., it assumes that with $30\%$, the whole probability support of transition, the agent will jump to the spot with the highest cost-to-go. The first distributionally robust approach takes into account an additional information by using two {\em nested} confidence sets: the jump probability parameter belonging to $9\%-11\%$ is of a confidence $1-\lambda$. The second distributionally robust approach incorporate more information. In specific, we construct an extra confidence interval, which is {\em disjoint} with above $9\%-11\%$ interval, and the chance of jumping with probability $20\%$ is $\lambda$. 

\begin{figure}[!t]
	\centering
	\includegraphics[width=2.8in]{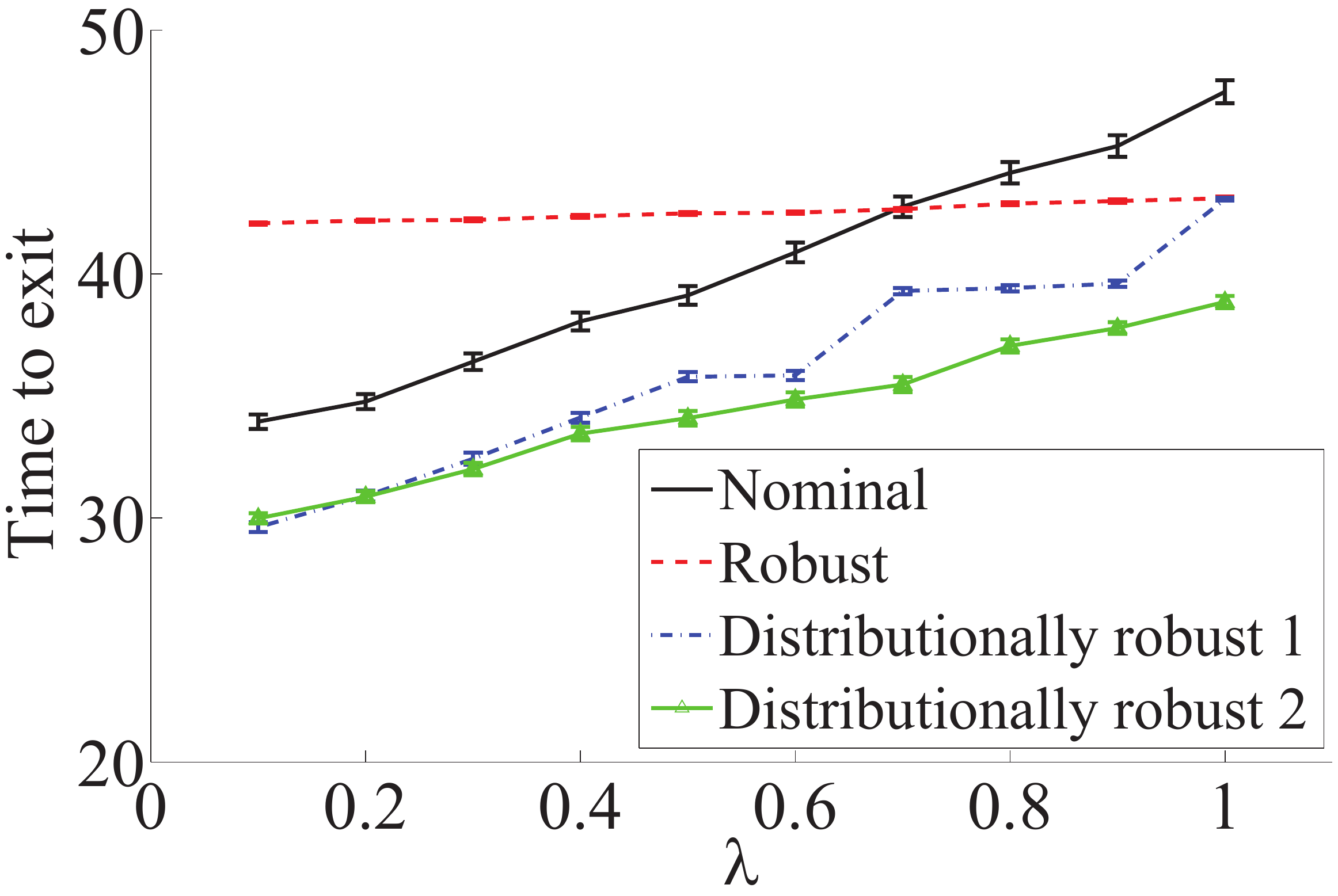}
	\caption{Performance comparisons between nominal, robust and two distributionally robust strategies on 3,000 runs of the path planning problem.}\label{figure8}
\end{figure}

The performance of strategies of the nominal, the robust and the two distributionally robust approaches is presented in Fig. \ref{figure8}, where the error bars show the standard error of the expected time to exit. The CPU times of computing optimal policies for four strategies are 0.461, 549, 642 and 654 seconds, respectively. The second distributionally robust approach, i.e., the approach proposed in this paper, outperforms the other three approaches over virtually the whole spectrum of $\lambda$. This is well expected, since additional probabilistic information is available to and incorporated by the second distributionally robust approach which considers ambiguity sets with more general structures.

\section{Conclusion}\label{sec6}

In this paper, we considered Markov decision problems with uncertainty. Specifically, we generalized the distributionally robust approach proposed in \cite{xu2012distributionally} to incorporate more general ambiguity sets \cite{wiesemann2013distributionally} to model a-priori probabilistic information of the uncertain parameters. We proposed a way to  compute the distributionally robust strategy through a Bellman type backward induction. We showed that the strategy, which achieves maximum expected utility under the worst admissible distributions of uncertain parameters, can be solved in polynomial time under some mild technical conditions. We believe that many important problems that are usually addressed using standard MDP models could be revisited and better resolved using the proposed models when parameter uncertainty exists, as this formulation naturally enables the decision maker to account for more general parameter uncertainty.

\ifCLASSOPTIONcaptionsoff
  \newpage
\fi



\bibliographystyle{IEEEtran}
\bibliography{reference}
\end{document}